\documentclass[12pt]{amsart}
\usepackage{a4wide,palatino,epsfig,latexsym,
natbib,amsmath,amsthm,graphicx,hyperref,enumerate,morefloats,color}
\usepackage{verbatim,caption}

\begin{document}

\title{The Changing Geometry of a Fitness Landscape Along an Adaptive Walk}  
\author{Devin Greene and Kristina Crona \\ 
University of California, Merced, CA USA}


\maketitle

\begin{abstract}
It has recently been noted that the relative prevalence of the various 
kinds of epistasis varies along an adaptive walk.  This has been 
explained as a result of mean regression in NK model fitness landscapes.  
Here we show that this phenomenon occurs quite generally in fitness 
landscapes.  We propose a simple and general explanation for this phenomenon, 
confirming the role of mean regression.
We provide support for this explanation with simulations, and discuss 
the empirical relevance of our findings.  
\end{abstract}


Corresponding author: Kristina Crona, kcrona@ucmerced.edu

\section*{Author Summary}
The main result concerns the changing geometry along an
adaptive walk in a fitness landscape.
An adaptive walk is described by a sequence of 
genotypes of increasing fitness, where two 
consecutive genotypes 
differ by a point mutation.
We compare patterns of
epistasis, or gene interactions,
along adaptive walks.
Roughly,
epistasis is antagonistic
(rather than synergistic)
if the double mutant combining
two beneficial mutations
has lower fitness than
expected. In the extreme case
that the double mutant has lower fitness
than one (or both) of
the single mutants,
one has sign 
epistasis. 
We claim that 
the further one is along an adaptive walk, 
the larger the frequency of sign epistasis 
and the smaller the relative amount of antagonistic epistasis relative 
to synergistic epistasis. 
We provide a simple and general
argument for our claim,
which hence likely applies to
empirical fitness landscapes. Our
claims can readily be checked 
by empirical biologists.
Potential theoretical progress
related to our work includes
a better understanding of
the role of  recombination
in evolution.

\section*{introduction}
Darwinian evolution can be illustrated
as an uphill or adaptive walk in a multidimensional landscape,
where one dimension (height) corresponds to genotype fitness, and the 
geometry of the remaining dimensions is determined by the locus--wise
 mutational distances between the genotypes.   
The metaphor
of a fitness landscape 
was introduced by \citep{w},
and has been formalized in
various ways, see e.g.\  \citet{bpse}
for a discussion. 
The fitness landscapes
we consider here are called genotypic.
A very basic type of a fitness 
landscape is one where mutation at a locus 
has a uniform effect regardless of the
state of the other loci (or {\it background} in the usual parlance).  
In most models, this effect is either additive or multiplicative.   
Deviations from this basic type occur when the effect on fitness 
of a mutation at a particular locus is dependent of the state of 
the other loci.  The general term for such background dependence 
is {\it epistasis}.  
We study how epistasis varies along an adaptive walk
in a fitness landscape. The topic is important
for understanding how a population adapts
after a recent change in the environment.
Several empirical 
studies \citep[e.g.][]{ccd,kds}
suggest that  the adaptation process
changes character over time, and
the role of epistasis may be critical.
The description of the changing form of epistasis
given in \cite{dp} is the starting point
for this work.  

To simplify our discussion, we will restrict ourselves to the 
following model.  A fitness landscape consists of all possible 
genotypes with a finite number of loci, denoted $L$, each biallelic, together with the fitnesses
of the genotypes.  
In this manner, we have a one--to--one correspondence between the set of 
possible genotypes and the set of bit strings of length $L$.  Fitnesses of 
genotypes are taken to be multiplicative, in the sense that the ratio of 
fitnesses of one genotype compared to another is the relative 
reproductive success of the fitter compared to the less fit.  
In this study, epistasis will be a feature associated with a quadruple 
of genotypes which differ by at most two loci.  
When considering such quadruples we will denote one genotype as a base, 
$ab$, two single mutants $Ab$ and $aB$, and the double mutant $AB$.  
If it is assumed that $ab$ has lowest fitness of the four, we can 
represent the fitness relations among the four genotypes 
by the graphs shown in Figure \ref{4squares}.

\begin{minipage}[h]{\linewidth}
\includegraphics[width=\linewidth]{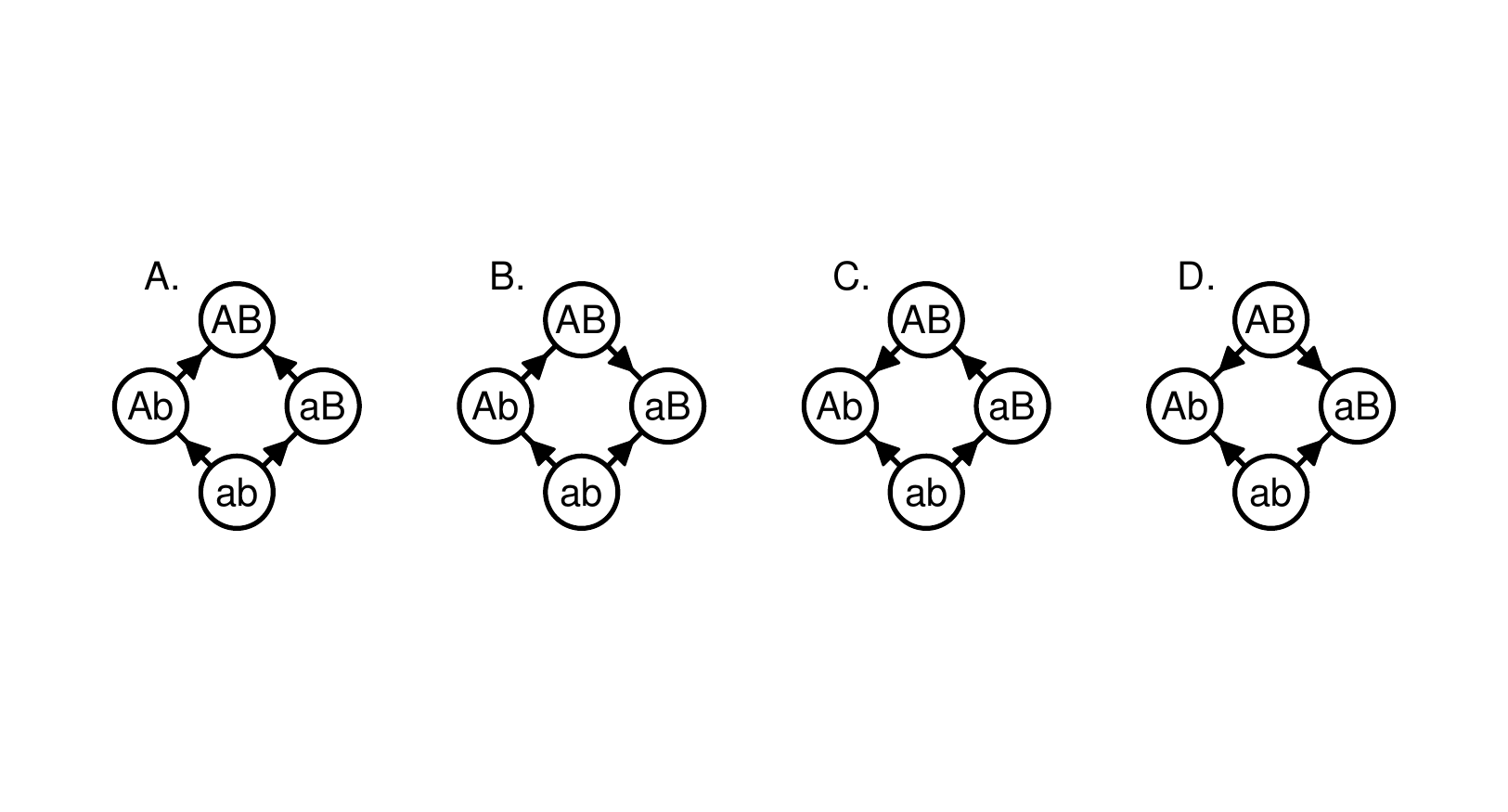}
\captionof{figure}{\small \it Two biallelic loci corresponds to four genotypes.
The possible fitness relations between neighbors
are illustrated in the graphs, where
each arrow points toward the 
genotype with higher fitness.}
\label{4squares}
\end{minipage}

\bigskip

Fitness graphs provided an intuitive way of 
representing a fitness landscape or its parts.  The vertices of the 
fitness graph represent genotypes.  
Arrows connect mutational 
neighbors, with the arrow 
pointing toward the genotype of higher fitness.  
An adaptive walk can then be 
viewed as a path in the 
graph respecting the direction of the arrows.  
Fitness graphs have been used
for displaying empirical data \cite[e.g.][]{ dpk, fkd},
and for deriving theoretical results \citep{cgb, c}.

Cases B, C, and D in Figure \ref{4squares}
present a situation where a mutation at one locus changes the 
direction of the fitness effect of a mutation at the other locus.  
Quadruples of genotypes which exhibit one of these relationships 
are said to exhibit {\it sign} epistasis,
a widely used concept first introduced in \citet{wwc}.
For more background relevant in this 
context, see 
\citet[e.g.][]{pkw, psk, cgb, c}.
Several studies of
empirical fitness landscapes
concern antimicrobial drug resistance,
where sign epistasis  seems
to occur for most landscapes where $L\geq 4$ 
(see e.g. \citet{ssf} for a survey of empirical
fitness landscapes.)

The type of non--sign epistasis in case A of Figure \ref{4squares}
is determined by the sign of the quantity 
$D = w_{AB}w_{ab} - w_{Ab}w_{aB}$, where $w_{ij}$ is 
the fitness of the genotype $ij$.  
When $D$ is positive, the quadruple is said to
have {\it synergistic} epistasis, when negative, 
{\it antagonistic} epistasis.  Conceptually, synergistic epistatis occurs 
when genotype $AB$ has superior fitness to what would be expected under 
a multiplicative model based on the fitnesses of $ab$, $Ab$, and $aB$, 
while antagonistic epistasis occurs when $AB$ has inferior 
fitness to what would be expected.  
Throughout the paper, we will restrict the descriptions synergistic 
and antagonistic to non--sign epistasis.

In \citet{dp} it was found that the prevalence of the three categories 
of epistasis undergoes significant change along an adaptive walk, 
with sign epistasis increasing in frequency as the walk progresses, 
and antagonistic epistasis decreasing relative to sign epistasis and 
marginally decreasing relative to synergistic epistasis.  
The authors discuss the phenomenon in some generality and analyze
empirical examples. 
However, in their explanation, the authors confine themselves to NK models
\citep[e.g.][]{kl,kw}, 
and their arguments are dependent of the details of how 
NK models are defined and constructed.  

The goal of this study is to 
investigate this phenonomen among a more general class of 
fitness landscapes, and provide an explanation  
independent of model specific assumptions.    
We appreciate that the classical models, including the NK model
are valuable for testing ideas. However,  explanations
independent of structural assumptions on
the landscapes are desirable,
especially since it is unclear how relevant the classical 
models are for empirical fitness landscapes.

\begin{figure}
\begin{center} 
\includegraphics[width=2in]{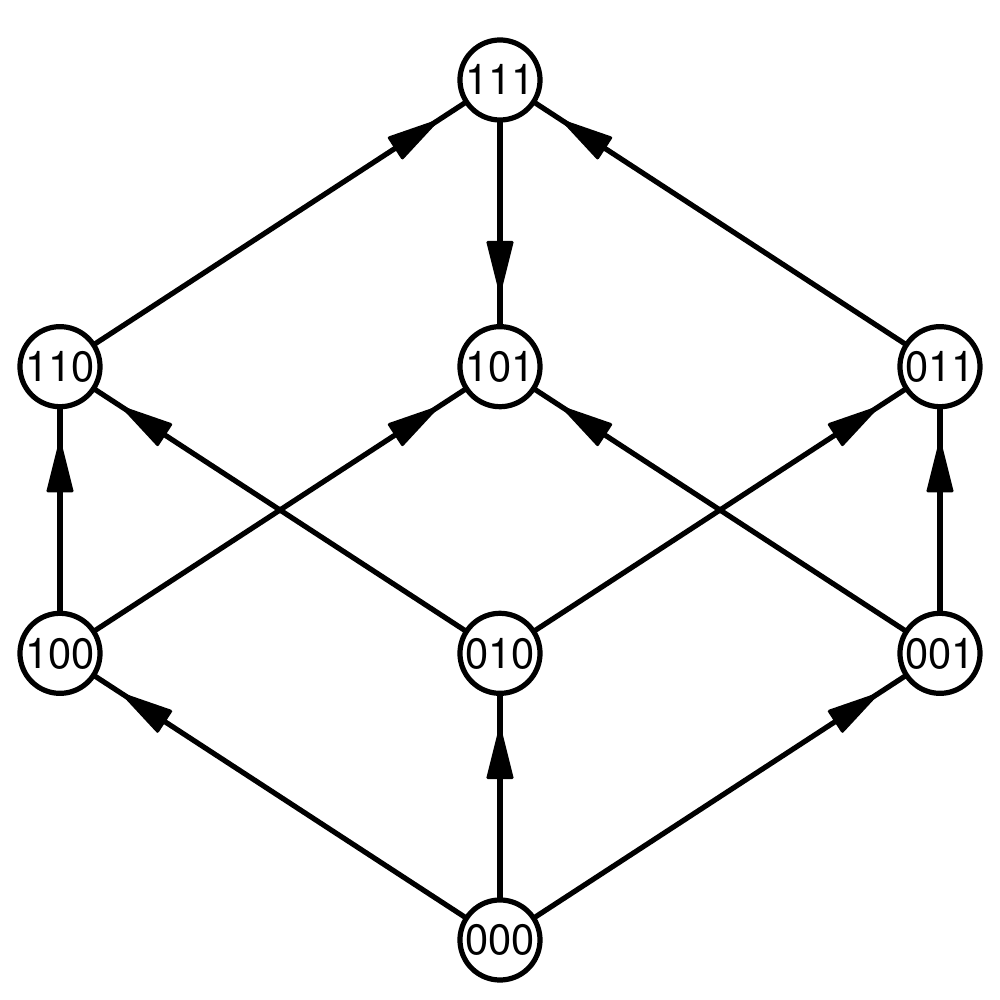}
\end{center}
\caption{\it \small A fitness graph for three loci.
}
\end{figure}

\section*{Results}

We consider two types of fitness landscapes in our simulations: 
NK models and ``Rough Mt.\ Fuji'' models 
\citep[e.g.][]{ah,aui,fkd}.  
The precise definition of both types 
of landscapes are found in the Appendix.  
Briefly, the fitnesses of genotypes in an NK landscape are 
determined by the fitness contribution of each locus.  
The fitness contribution of each locus is a stochastic function of its 
own state plus the state of K other loci which are fixed in advance.  
When K = 0, the landscape is purely multiplicative 
(or additive, depending on our choice of model), and 
(in the multiplicative case) would have no epistasis.  At the other extreme, 
when K=N-1, the fitnesses of genotypes are mutually independent, leading to 
abundant epistasis.  

The so called Rough Mt.\ Fuji models  
are constructed by starting with a purely 
additive or multiplicative model, where each allele contributes a fixed, equal
amount, independent of background.  The determinate fitnesses obtained this way are then 
perturbed by random noise.  See the appendix for further details on the construction of 
Rough Mt.\ Fuji landscapes.
In this study we confine ourselves to additive Rough Mt.\ Fuji landscapes, 
though we note that simulations performed with multiplicative 
Rough Mt.\ Fuji models (and which are not reported in this study) support the conclusions below.
We fine tune the relative magnitudes of 
random noise and fixed additive contribution with a parameter, 
thereby allowing us to vary Rough Mt.\ Fuji landscapes in a manner 
analogous to varying NK models with the choice of K.  

We will be concerned with the properties of adaptive walks in 
our fitness landscapes.  We will assumes the 
asymptotic condition of Strong--Selection--Weak--Mutation (SSWM for short) 
 \citep{g1983, g1984, s}.  
It is assumed that the evolving population remains genetically 
monomorphic outside of very short time intervals, during which 
a new beneficial mutation sweeps to fixation.  Given a 
genotype $g_0$, population genetics theory shows that if the 
selection coefficients of the fitter mutational neighbors 
$g_1,\, g_2,\, \ldots,\, g_L$ of 
$g_0$  are $s_1,\, s_2,\, \ldots,\, s_L$, respectively, then the 
probability of $g_i$ going to fixation is 

\[
\frac{s_i}{\sum_{i=1}^n \, s_i}.
\]
\noindent (It should be noted that we are sweeping under the rug the fact that 
strictly speaking this formula is appropriate only when the magnitudes of the second or higher powers of the $s_i$ are negligible.)
For more background about the SSWM assumption, as 
well as the fixation probability described, see
\citet{o}.

An adaptive walk, then, can be viewed as a stochastic path 
in a fitness landscape, starting at an initial genotype 
and ending at a genotype with locally maximal fitness.  
For every two steps in such a walk, three genotypes are 
traversed, which can be denoted, in order, $ab$, $Ab$ ,and $AB$.  (Note that we are no longer assuming
the minimality of $ab$ as was done in Figure \ref{4squares}.)  
These genotypes are complemented by $aB$, and the type and magnitude 
of epistasis for the quadruple can be determined by their fitnesses.  
Note that the configuration in Figure \ref{4squares} \,D has no relevance for adaptive walks, and makes no
appearance in subsequent calculations.

In \citet{dp}, it was noted that the relative frequencies of sign, 
antagonistic, and 
synergistic epistasis varied along adaptive walks.  
Our aim is to explore this phenomenon more closely.  
What are the relative frequencies of sign, antagonistic, and synergistic epistasis?  

In our notation, we assume that three genotypes ab, AB and AB are traversed
in some adaptive walk, so that 
\[
w_{ab}<w_{Ab}<w_{AB},
\]
and consequently $w_{aB}$ determines the type of epistasis
(again, we do not assume that $w_{ab}$ is minimal). These
assumptions hold for the remainder of this paper.
The possibilities are that $w_{aB}$ is ranked first, 
second, third or fourth in terms of fitness relative to the other three 
genotypes. {\emph{When ranked first or fourth, the quadruple has sign epistasis, 
and not so when ranked second or third.}}
This fact will be used repeatedly.

We start with a preliminary observation.
In the special case where fitnesses of mutational neighbors are 
identically and independently distributed, such as in an NK landscape 
with $K = N-1$, 
and where the genotypes are chosen randomly,
the probabilities
that $w_{aB}$ is ranked first, 
second, third or fourth
are readily calculated.
Indeed, the probabilities
are equal, since the fitness of a
paticular genotype
is independent of mutational 
neighbors. Consequently
sign epistasis occurs 
with frequency $0.5$.

Similarly, consider a randomly chosen quadruple but
in landscapes where the fitness of mutational neighbors are correlated, 
as in NK landscapes with $K < N - 1$. Then we expect the frequency of sign epistasis 
to decrease relative to the case of uncorrelated fitness.  This expectation is confirmed by simulations, 
the results of which are found in the Appendix.  The parameter ${\it slope}$ in the 
Rough Mt.\ Fuji models is positively associated with correlation between 
mutational neighbors.  (See Appendix.)  The simulation results thus confirm the expectation 
of lower sign epistasis in landscapes with correlated mutational neighbors.  

The results of our
 simulations confirm \citet{dp}, namely that the further one is along an adaptive walk, 
the larger the frequency of sign epistasis 
and the smaller the relative amount of antagonistic epistasis relative 
to synergistic epistasis.  Significantly, a similar evolution of 
relative frequencies occurs in the Rough Mt.\ Fuji landscapes.  It is clear 
that a more general explanation for this phenomenon is desirable, since 
Rough 
Mt.\ Fuji fitness landscapes are not defined in terms of locus--by--locus 
fitness contributions.

We hypothesize that the observed evolution of
epistasis along adaptive walks is merely the familiar statistical 
phenomenon of 
regression to the mean.
This explanation was suggested in \citet{dp} as well. However, the authors'
arguments are restricted to the details of the NK model.  We offer here a simpler and more general 
explanation.

We begin with an intuitive explanation for the phenomenon we seek to explain.  This will be followed by evidence from simulations that support our argument.
We consider the type of epistasis that would be found with respect to a quadruple of genotypes $ab$, $Ab$, $aB$, and $AB$, where
$ab$, $Ab$, and $AB$ form three subsequent genotypes in an adaptive walk.

Informally, the following extreme example will clarify
important mechanisms. Suppose that $ab$ belongs to the highest 
fitness percentile among genotypes
in the fitness landscape. For uncorrelated fitness,
the expected frequency of
sign epistasis would be at least 99 percent. 
Indeed, one would get $w_{aB}<w_{ab}$ in 99 percent of
the cases.  Similarly, for correlated fitness   
one would many times get $w_{aB}<w_{ab}$ as well,
provided there is sufficiently much noise in
the landscape. This is because
a mean regression effect
will tend 
 to "pull" $aB$ below $w_{ab}$, since
$ab$ belongs to the highest fitness percentile.


After the informal example,
we now go over the different possibilities for the quadruple of genotypes in some detail.
We will compare low and high fitness of $ab$ with the "null" condition where $ab$ is randomly chosen.
If we impose the condition that $ab$ has lower fitness relative to the mean fitness of the landscape, then it is likely that $Ab$ and $AB$ will have lower fitness 
than would have been expected if $ab$ had been randomly chosen (unless the fitness landscape 
is uncorrelated, of course), though the likelihood of large jumps in the adaptive walk may return $AB$ to more typical fitness levels.  
To the extent $w_{aB}$ is determined by a stochastic component independent of $w_{ab}$, $w_{Ab}$, and $w_{AB}$, mean regression implies that it is more likely 
that $w_{ab} < w_{aB}$ than in the case where $ab$ is randomly chosen without condition from the fitness landscape. 
Note that the imposed condition of relatively low $w_{ab}$ biases the probability toward non-sign epistasis relative to the ``null'' condition.
Furthermore, within the region of non--sign epistasis, the bias toward 
$w_{aB} > w_{ab}$ relative in the null situation results in a higher probability
that $D = w_{AB}w_{ab} - w_{aB}w_{Ab}$ is negative, leading to a bias toward antagonistic epistasis. 
(One may ask about the possibility of sign epistasis
where $w_{aB}>w_{AB}$. However, the fact that $w_{ab}$ is low,
does not necessarily mean that $w_{AB}$ is low, as explained. 
In summary, we have all reasons to believe that
the decrease of cases of sign epistasis where $w_{aB}<w_{ab}$
outweighs a possible increase of sign epistasis where $w_{aB}>w_{AB}$,
relative the "null" condition.)

Conversely, when an adaptive walk reaches $ab$ after
a number of steps, and continues to $Ab$ followed by $AB$, it is highly likely $ab$, $Ab$, and $AB$ have high fitness relative to the mean fitness of the fitness landscape. 
To the extent that $w_{aB}$ is determined by a stochastic component independent of $w_{ab}$, $w_{Ab}$, and $w_{AB}$, mean regression implies that $w_{aB} < w_{ab}$ 
is more likely than would be the case when $ab$ is randomly chosen without condition.
Furthermore, within the interval of non--sign epistasis, the quantity 
$ D = w_{AB}w_{ab} - w_{aB}w_{Ab}$
 is biased upward toward positive values, thus leading to 
a higher proportion of synergistic epistasis to antagonistc epistasis.  
We conclude that the changing balance of types of epistasis along an 
adaptive walk is not due to any intrinsic feature of adaptive 
walks per se, but rather the result of traversing from lower to higher fitnesses.  Late stage adaptive walks are ``walking along a ridge'', implying more sign epistasis.

In summary, the pattern of changing epistasis along an adaptive walk is driven by mean regression due to the fitnesses of $ab$, $Ab$, and $AB$ and the uncorrelated component of the fitness of $aB$.  

Finally, we stress that the observed phenomenon relies on
an important asymmetry between
$w_{ab}$ being far below the mean, and far above the mean.
Indeed, the quantity $|w_{AB}-w_{ab}|$ is expected to be 
relatively large for very low $w_{ab}$, and relatively small
for very high $w_{ab}$. In particular, this asymmetry helps explain why one
expects a prevalence of sign epistasis when $ab$ has high fitness.

Figure \ref{walks} depicts 
the patterns of epistasis along adaptive
walks. The patterns agree well with our intuitive 
description. The figure concerns the NK landscape
with parameters $N=15$ and $K=1$.  See the
Appendix for a complete description
of our simulations of adaptive walks.

\begin{minipage}[h]{\linewidth} 
\begin{center}
\includegraphics[width=\linewidth]{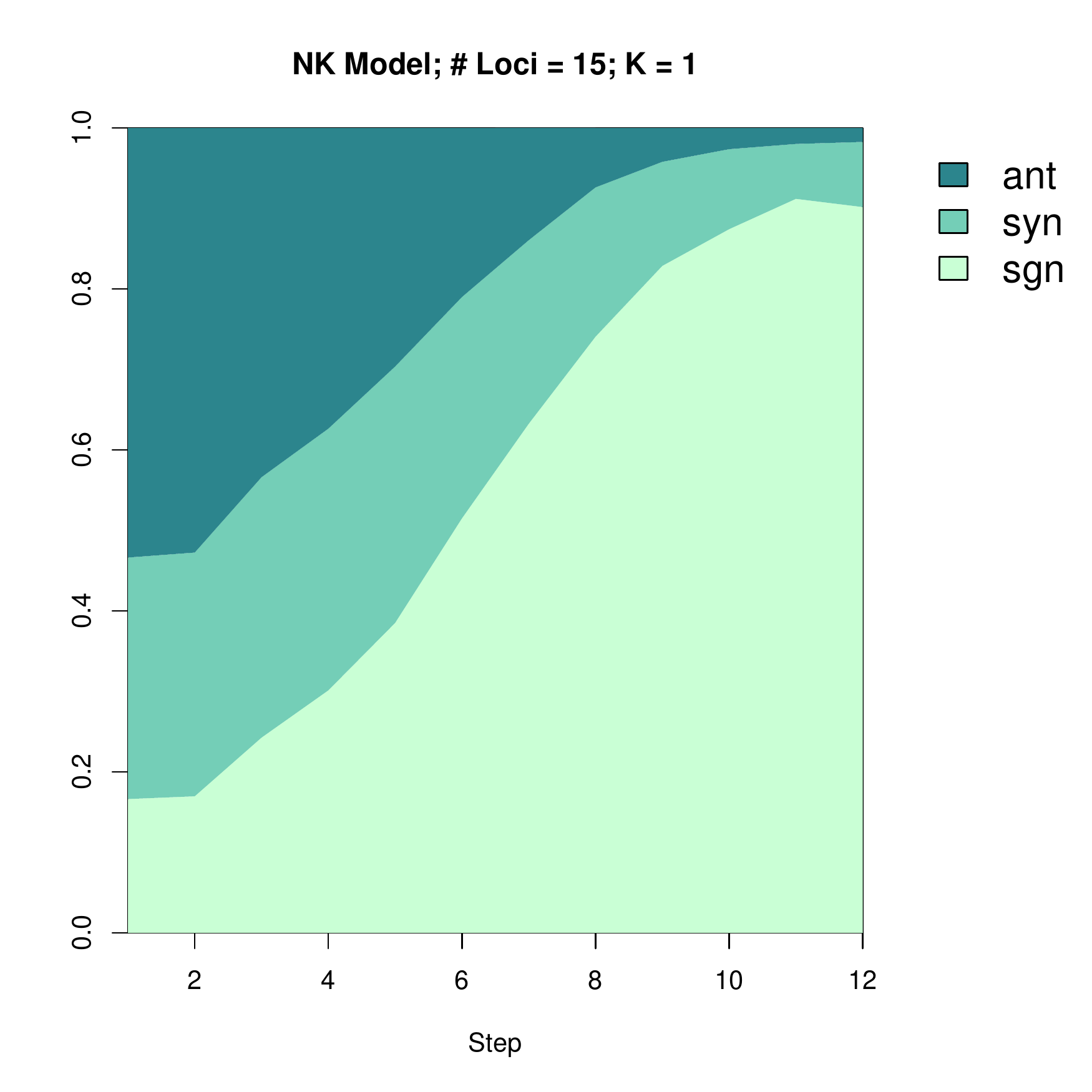}
\end{center}
\label{walks}
\captionof{figure}
{\small \it 
According our simulations, the patterns of 
epistasis change along adaptive walks as
displayed. The graph depicts NK landscapes
with parameters $N=15$ and $K=1$. }
\label{walks}
\end{minipage}

The case of high$w_{ab}$
 is illustrated somewhat crudely in Figure \ref{star}.
The blue arrows form part of an adaptive walk, and the three vertices they connect correspond to $ab$, $Ab$, and $AB$ 
above.  If we assume that $ab$ has higher than average fitness, then when the fitness of genotype $aB$ has an uncorrelated component there is a bias toward $w_{ab} > w_{aB}$, leading to sign epistasis.

We buttressed our intuitive argument above by examining the results of simulated fitness landscapes and adaptive walks. The results of these simulations are attached as a supplement to this 
article.  If our explanation above is correct, two results should
emerge from our simulations.  
One, if random quadruples of genotypes as shown in Figure \ref{4squares} are sampled in a stratified fashion from different fitness quartiles of the landscape, 
then the frequencies of sign, antagonistic, and synergistic epistasis should change their relative proportions from the lowest quartile to the highest quartile
as they do along an adaptive walk.  They do, as 
can be seen in Figure \ref{bars} and in the Appendix.
(To clarify, we sampled $ab$ so that $w_{ab}$ belongs to the specified quartile.
We did not impose any condition on $Ab$ and $AB$, except that
$w_{ab}<w_{Ab}<w_{AB}$).

Two, if we simulate adaptive walks under the condition of equal probabilities among 
all mutational neighbors, the rate at which fitness increases should be slowed, and therefore the 
frequencies of types of epistasis should change at a slower pace than they do in a weighted probability model.  They do, as can be discerned by comparing 
the figures with equally weighted probabilities, to the figures with probabilities weighted according to the SSWM model (see the Appendix).

\begin{minipage}[h]{\linewidth}
\includegraphics[width=\linewidth]{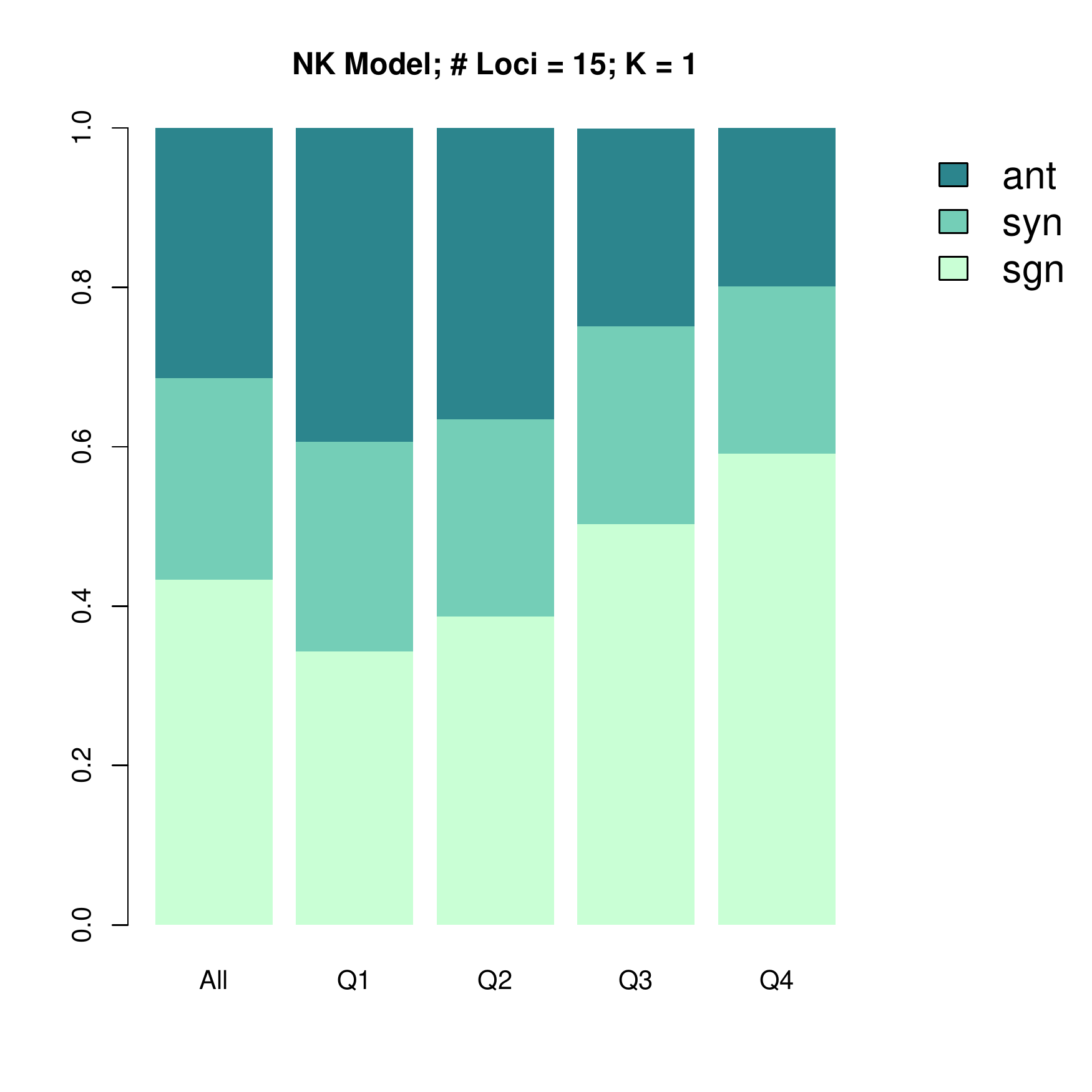}
\captionof{figure}
{\small \it 
Random quadruples were sampled in a 
stratified fashion,  where $w_{ab}$ 
belongs to the specified fitness quartile.
The frequencies of sign, antagonistic, and synergistic epistasis should change their relative proportions from the lowest quartile to the highest quartile
as they do along an adaptive walk.}
\label{bars}
\end{minipage}


Further support for our proposed explanation was obtained by simulating 1000 $NK$ landscapes with $N=15$ and $K=10$. The result, summarized in   
Figure \ref{Confidence Intervals}, confirm our assertions.
  
For each landscape, a 
genotype with relatively low fitness was chosen as the initial genotype of an adaptive walk (see Appendix for details).  Figure \ref{Confidence Intervals} summarizes
the important features of the results of the simulations.  In caption A, $2.5\%-97.5\%$ percentile intervals are shown for the first($ab$), second($Ab$), and third($AB$) genotype of 
the adaptive walk.  The fourth interval corresponds to the complementary genotype $aB$.  The ranges of the intervals show a bias toward non-sign epistasis.  
The blue "control" interval corresponds to randomly selected genotypes.

Conversely, in caption B, $2.5\%-97.5\%$ percentile intervals are shown for the fourth($ab$), fifth($Ab$), and sixth($AB$) genotypes visited on an adaptive walk.  Again, the fourth interval
corresponds to $aB$.  In this case, the bias is toward high frequency of sign epistasis.

In both cases, the role of mean regression in driving the nature of epistasis along adaptive walks is apparent.  
s \ref{fitnesslandscape1} and \ref{fitnesslandscape2} represent partial views 
of one simulation as described above.  Even here, the bias toward or away from sign epistasis depending on the stage of the adaptive walk is apparent.

We have compared equal weights,
and adaptive walks under the SSWM assumption.
For more background and results regarding
lengths of walks,
we refer to \citet{mp,fl}
for equal weights, and 
\citet{o} for the SSWM case.

As a final remark,  the study of epistasis as described was restricted to pairwise interactions.  It would be interesting
to extend the study to higher order interaction, and for instance to consider shapes as defined in the geometric theory of
gene interactions \citep{bps, bpse}.
\vspace{0.0in}

\begin{minipage}[b]{\linewidth}
\begin{center}
\includegraphics[width=2.4in]{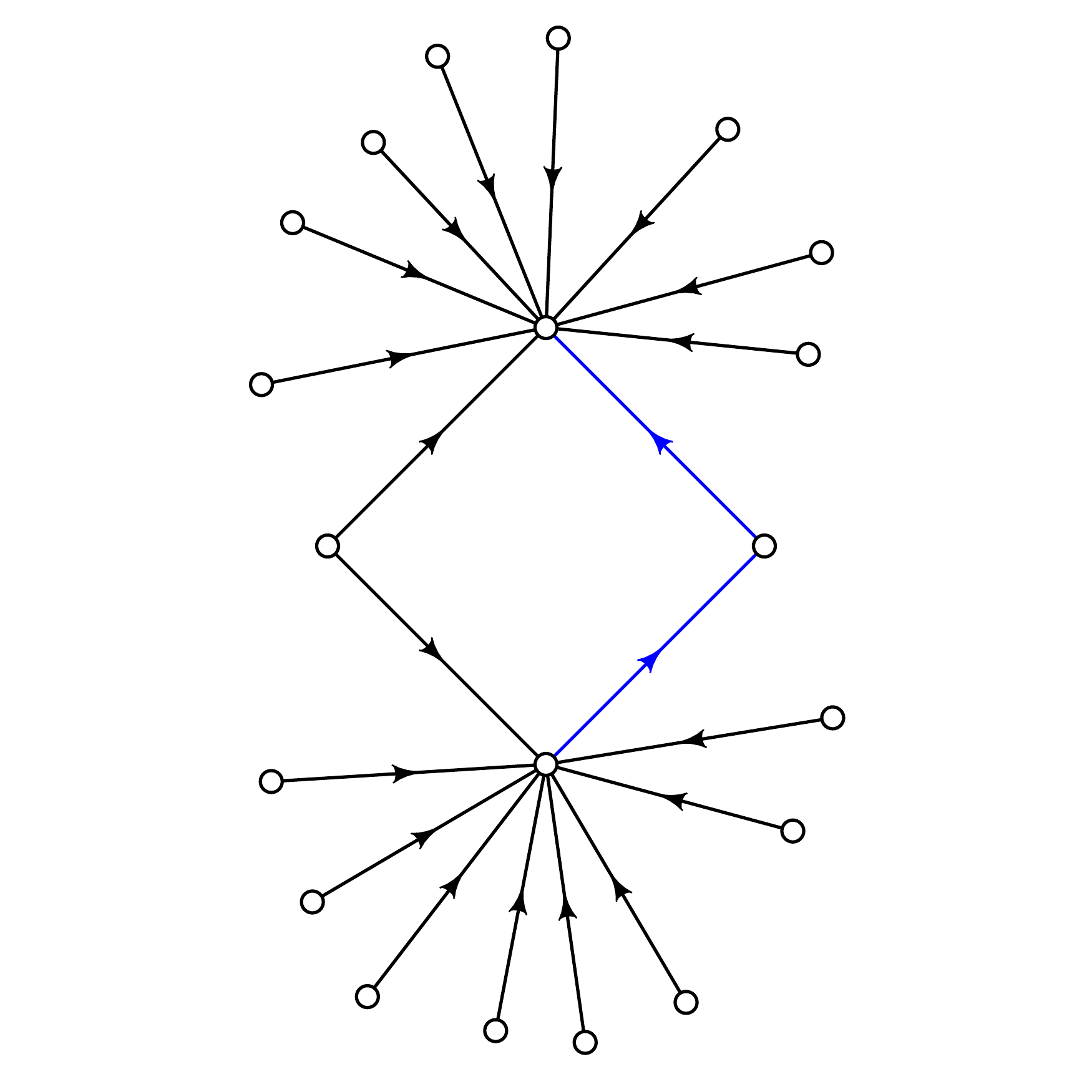}
\captionof{figure}{\small \it Assume that the
adaptive steps, colored blue,
connect three genotypes 
with relavatively high fitness.
Most connecting arrows point toward the
starting point, as well as the end point of
the adaptive steps. Note that due to the high fitness of the
genotypes along the adaptive walk, the arrows emanating from 
the fourth genotype in the quadruple are more likely to point outward.  
The result in such a case is sign epistais.  }
\label{star}
\end{center}
\end{minipage} 

\begin{minipage}[h]{\linewidth}
\begin{center}
\includegraphics[width=6in]{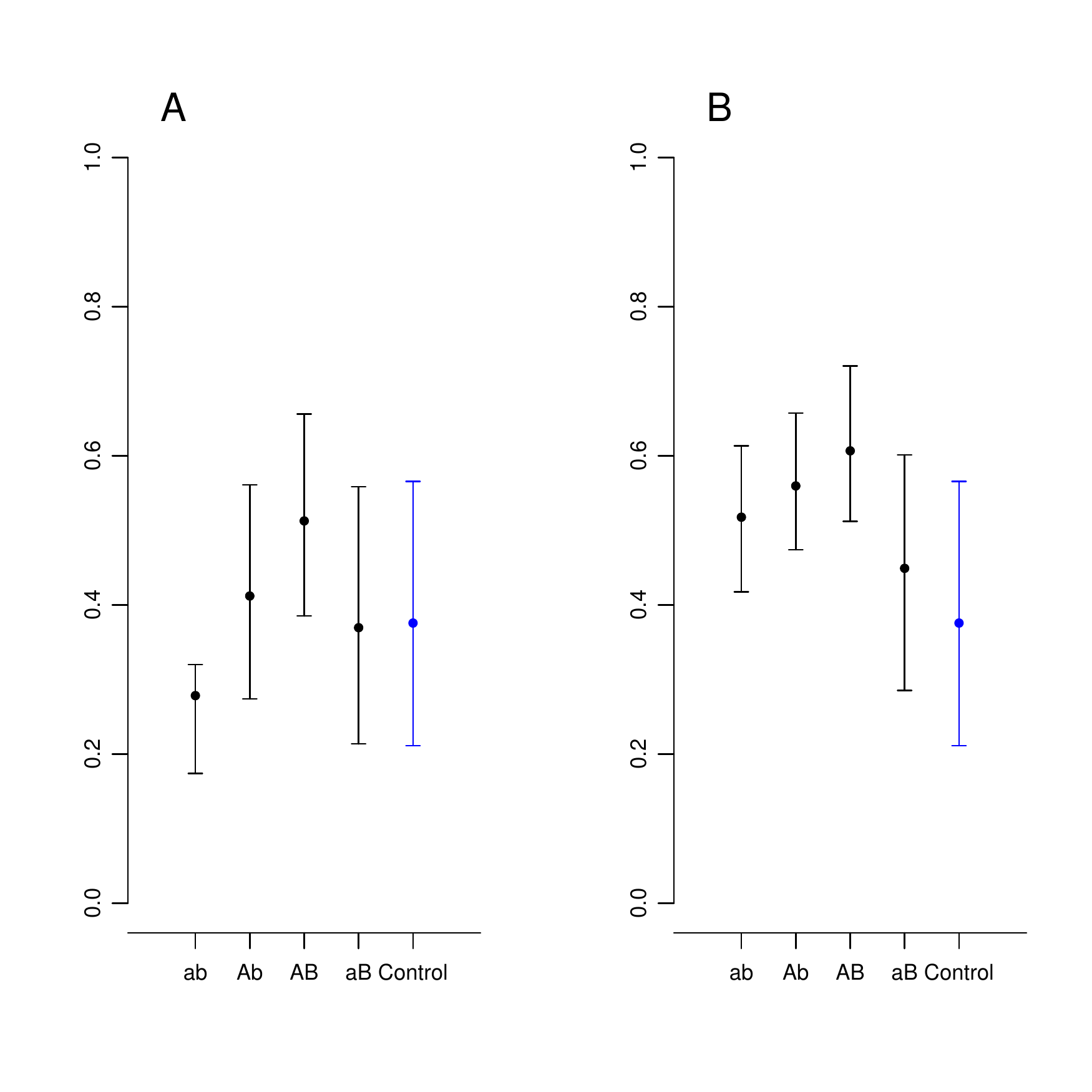}
\end{center}
\captionof{figure}{ \small \it
1000 adaptive walks simulated on NK landscapes with N=15 and K=10.   
For each walk, the starting genotype $ab$ was randomly drawn to have relatively low fitness (see Appendix for details).  {\bf A.} Intervals covering fitnesses between the 2.5 and the 97.5 
percentiles are shown for the first (ab), second ($Ab$), and third ($AB$) genotypes in randomly generated adaptive walks, with dots indicating the medians.  The genotype $aB$ is the remaining 
genotype in the quadruple as shown in  
Figure \ref{4squares}.  The blue ``Control'' interval corresponds to randomly selected genotypes.
The skew visible in the ab interval is due to the fact that the initial genotype of a fitness walk is drawn from a lower tail distribution.  {\bf B.} Intervals for the fourth, fifth, and sixth genotypes in 
randomly generated adaptive walks.  The increased fitness of the aB genotypes in B relative to that of A is due to the fact that $K=10 < 14$, and thus there is some correlation between 
neighboring genotypes.  In both diagrams, the dependency of sign epistasis on regression to the mean is apparent.  
}
\label{Confidence Intervals}
\end{minipage}

\section*{\scriptsize Empirical support and applications}
As mentioned in the introduction, empirical data seem to support the 
``mean regression'' hypothesis exposited herein.  We add further support with the following 
empirical results from investigations of the TEM-family of $\beta$-lactamases \citep{gmc}.
The TEM-enzymes are associated
with resistance to several
$\beta$-lactame antibiotics, 
including  penicillins. 
TEM beta-lactamases have been found in Escherichia coli, Klebsiella
pneumoniae and other Gram-negative bacteria.
TEM-1 is
considered the wild-type,
and approximately 200 mutant variants  
have been found clinically,
(see e.g.\ the record 
from the Lahey Clinic \url{http://www.lahey.org/Studies/temtable.asp}).

For the 4-tuple mutant TEM-85 (L15F, R164S, E240K, T265M)
the two fitness landscapes
defined by Cefotaxime and Ceftazidime
had mutational trajectories (i.e.\ adapative walks) from
TEM-1 to TEM-85. For Cefotaxime there were three 
trajectories to TEM-85, and for Ceftazidime
one trajectory. We calculated the 
epistasis in the last two steps, as well
as in the first two steps, of the four trajectories.
Fitness differences of mutational
neighbors were not always statistically significant
in the study, 
resulting in cases of ``possible'' sign epistasis.
The results for the last two steps
were two cases of sign epistasis,
and two cases of possible sign
epistasis.
The results for the first two steps
were two cases of possible sign epistasis,
and two cases of no epistasis.
These findings seem to support our hypothesis, though we must refrain from drawing
any sweeping conclusions based on a small data set.

Generally speaking, there
are two types
of empirical studies of
evolution, direct and indirect.  
A direct study
is concerned with an evolving
population,
where mutations are
observed as they occur.
Examples of this are a population evolved in a laboratory  
or the stages of an HIV infection due to drug resistance conferring mutations.  
The second type 
of study is indirect. 
An investigator attempts to create a 
catalog of genotypes with the potential of being part of an adaptive walk.  
As an example, a strain of bacteria that is 
highly resistant to a particular antibiotic treatment may differ 
from the wild-type by $n$ amino acid substitutions in a relevant 
enzyme.  The investigator in an indirect study will attempt to 
produce and study all $2^n-2$ intermediate mutational stages.  
It is non-trivial to relate
direct and indirect studies.
One wishes to infer
the fitness landscape 
from an evolving population.
Conversely, one would like to predict evolution 
from indirect studies.
As observed in \cite{dp},
epistasis may influence
path choice for evolving populations,
and path choice has an impact on
epistasis. Consequently, 
it may be difficult to infer the fitness landscape from
a direct study.

As for the converse,
it may seem
straightforward 
to predict evolution
from a fitness landscape.
However, a practical difficulty arises; namely,
the information one has in an 
indirect study is often restricted to the fitness \emph{rankings} of the 
genotypes, with no quantitative measurements of 
fitness. Consequently, one
has very little knowledge of the probabilities
of evolutionary trajectories,
even if the fitness graph is known.

At issue here is the fact that examining epistasis in 
fitness graphs and evolving populations may lead to results 
which seem at odds. 
It is \emph{a priori} not clear if 
patterns of epistasis along adaptive walks
are easily predicted from fitness graphs. 
In addition to being used for confirming the 
robusticity of our results, we included the  equally weighted
adaptive walks (see the Appendix) to reflect the point of view of
the results of an indirect study, where only the fitness rankings of the genotypes 
in the landscape are discovered, and thus there is no \emph{a priori} knowledge 
of the appropriate weights to be assigned to the various paths evolution may follow.  

The pattern of epistasis was broadly held across the two 
classes of fitness landscapes considered here, across 
a range of parameters for these landscapes, and across the 
weighted versus the unweighted versions discussed above.
(The main
difference we could find 
was pace in which proportions of epistasis changed,   
which is easily explained by the fact that the rate of 
fitness increase is slower in the equally weighted walk.)
If we consider the equally weighted case as corresponding 
to indirect studies, and the weighted case to direct studies, 
then it is interesting to note while the rate of change of the proportions
varies, the general pattern does not. 
Naturally it would be interesting
to further investigate the
relation between direct
and indirect studies of 
adaptation.

\section*{Discussion}
The nature of epistasis varies along an adaptive walk.
This observation has been made in
simulations, and has support in
some empirical studies.
We have argued that mean regression 
is a simple and general explanation for this
phenomenon.
We support this explanation with simulations 
carried out on two classes of fitness landscapes, with varying 
parameters.
While our simulations were restricted to two classes, our argument
should extend to any fitness landscape where 
genotypes vary to any degree independently to each other.


We considered two types of adaptives walks; those 
with probability weight corresponding to those used in the 
SSWM model, and those with equal probability weights.   
The similarity of the results suggests that  
the pattern of epistasis found along an adaptive walk is not
a result of any specific property of adaptive walks generated according to 
the SSWM model.  This result is also relevant for relating direct and indirect studies
as defined above.

Further support for
our assertion was obtained by sampling genotypic quadruples of 
mutational neighbors from 
simulated fitness landscapes at different fitness quartiles. The resulting 
pattern of increasing sign epistasis and decreasing antagonistic to 
synergistic ratio at higher fitnesses relative to lower fitnesses reinforces our
assertion that the same phenomenon seen along adaptive walks depends on mean 
regression, and does not depend on any intrinsic properties of adaptive walks per se.


It should be pointed out that confidence intervals and issues with statistical power were ignored in this article.
For each set of parameters, we simulated $10,\!000$ fitness landscapes with an adaptive walk.  
It can be seen from the figures in the Appendix that for most types of landscapes the number of adaptive walks which evolve to an $m$th genotype before hitting a local 
optimum decreases quite significantly with $m$ after approximately the four steps.  Naturally, the low number of adaptive walks which attain higher steps may raise concerns 
of statistical power.  Nevertheless, despite this possible shortcoming, we feel that the general pattern is clear enough.

Our main observation has
important consequences for
interpretations of empirical data.
Consider any fitness landscape where
there is a well defined wild-type, and
some beneficial single mutants.
For instance, the fitness landscape
may be associated with antimicrobial
drug resistance.
Some recent papers consider
prevalence of sign epistasis,
and related questions for such
landscapes, where the wild-type
is used as a starting point
(for a survey article, see \citet[e.g][]{ssf}

Our result demonstrate
that there are two factors
that influence the prevalence
of sign epistasis [even for a given
parametric model].
The first is the degree of
additivity in the landscape.
The second is the fitness of
the wild-type. 
Ideally, a study should therefore
estimate wild-type fitness
as well as additivity in the
landscape. Roughly,
one can estimate 
wild-type
fitness from the proportion
of single mutants which are more fit
than the wild-type among all
mutational neighbors of
the wild-type
(see e.g. \cite{kelly} for more 
comments).

We have argued that our main
observation holds for empirical 
fitness landscapes. 
Most aspects of adaptation
are sensitive for epistasis.
In particular, a serious 
analysis of recombination,
requires a fine-scaled
understanding of epistasis.

Finally, our study was restricted
to pairwise interactions.
It would be interesting
to extend the arguments given here
to higher order 
interactions among loci.






\begin{minipage}{\linewidth}
\makebox[\linewidth]{
\includegraphics[width=5.5in]{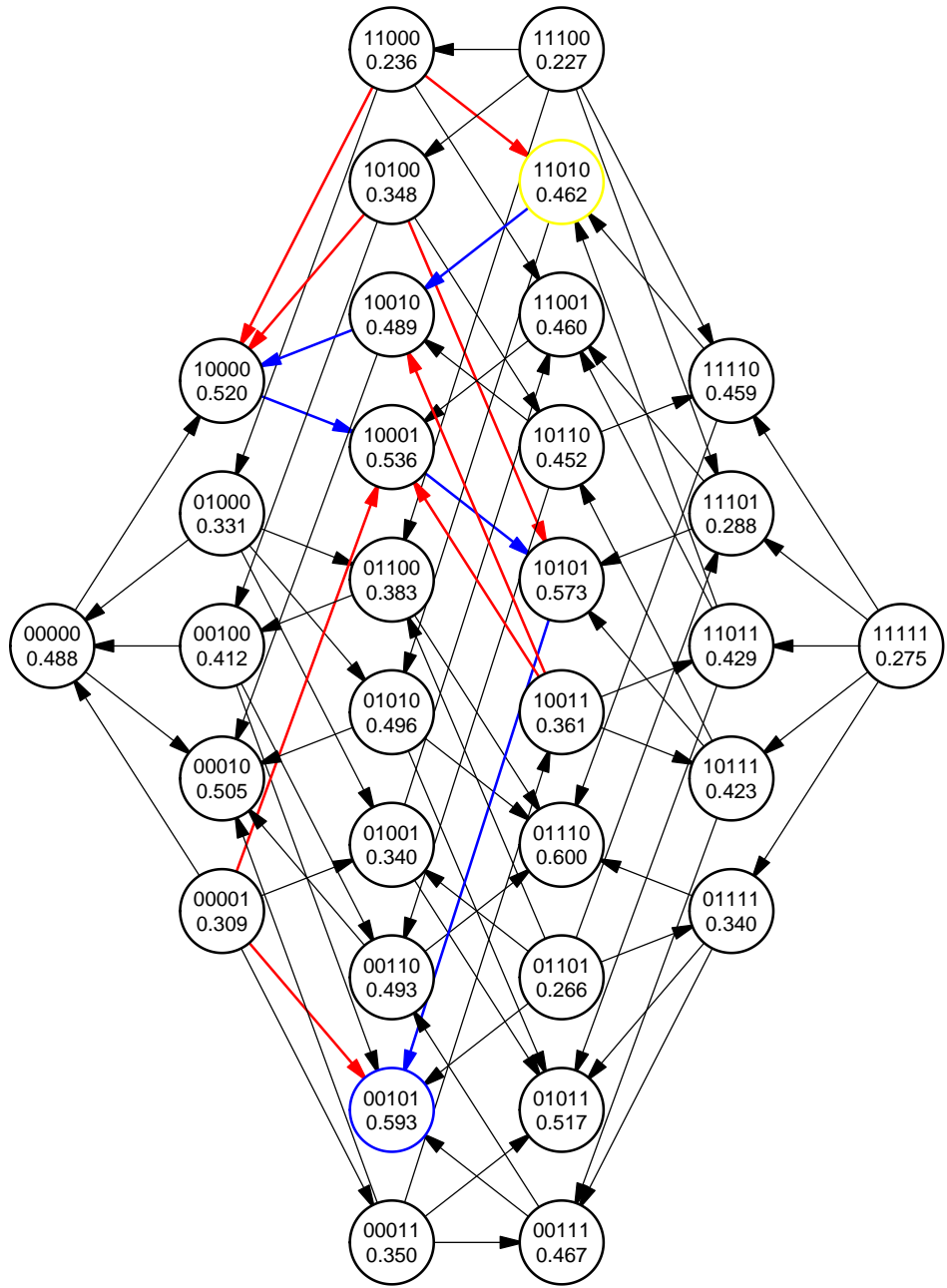}}
\captionof{figure}{\it \small A depiction of the fourth (yellow), fifth, sixth, seventh, and eighth genotype of an adaptive walk in an NK landscape, with $N=15$ and $K=10$.  
Only loci affected by mutation during the five adaptive steps are shown in the genotype labels,  
and the genotypes shown are restricted to those that differ from the initial genotype only at the five affected loci.  The fitness of each genotype is also shown.
The adaptive walk is colored blue, while the opposing arrows in each quadruple are colored red.  
Note the dominance of sign epistasis along the adaptive walk.  The ridge-like quality of the adaptive walk is clear from the high proportion of ``in'' arrows emanating from the evolved 
genotypes.
}
\label{fitnesslandscape1}
\end{minipage}

\newpage

\begin{minipage}{\linewidth}
\begin{center}
\makebox[\linewidth]{
\includegraphics[width=5.5in]{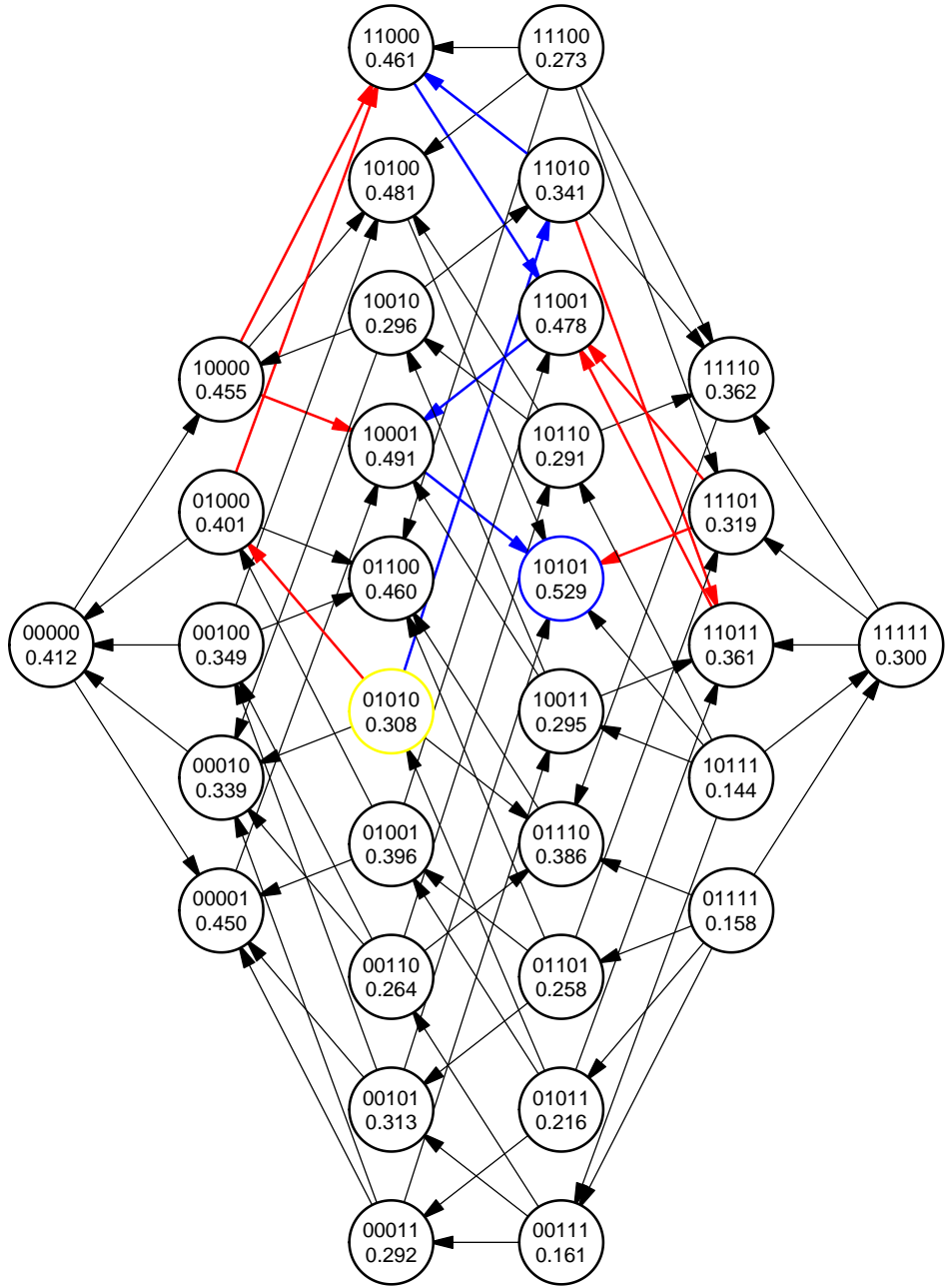}
}
\captionof{figure}{\it \small A depiction with a description analogous to Figure \ref{fitnesslandscape1}, but in contrast, the yellow colored genotype is the initial genotype of the adaptive walk.  Note the lower frequency of sign epistasis along the walk as compared to Figure \ref{fitnesslandscape1}.  }
\label{fitnesslandscape2}
\end{center}
\end{minipage}

\section*{Materials and Methods}

Throughout this study, loci were considered to be bi--allelic, with alleles $0$ and $1$ for each locus.  All of the fitness landscapes had 15 loci.

The NK model is classical.  The so--called Rough Mt.\ Fuji model has been explored. 

Some of the features of our fitness landscapes were peculiar for this study, so we will summarize briefly in this appendix how they were constructed.  

For the NK fitness landscapes, the contribution of each locus is a function of the allele at the locus itself as well as the alleles at $K$ randomly chosen additional loci, or 
w
$$w_j = w_j(l_j,l_{j_1},l_{j_2},\ldots,l_{j_K}), \quad j = 1,2,\ldots,N; \quad l_i = 0,1$$

The fitness of a particular genotype $l_1l_2\ldots l_N$ is then the geometric mean of the individual loci contributions:

\begin{equation} \label{bill} w(l_1l_2l_3\ldots l_N) = \left(\prod_{j=1}^N \, w_j(l_j,l_{j_1},l_{j_2},\ldots l_{j_K})\right)^{1/N}, \quad l_i = 0,1
\end{equation}

For each of the possible values of $w_j$, we sampled independently from a uniform distribution over the interval $[0.05,1]$.  The $0.05$ floor was used to prevent overly large fitness 
coefficients.  

Since calculating the fitness of each genotype in an NK landscape proved computationally time--consuming, we determined the fitness quartiles theoretically as follows.  Since the logarithm of 
the right hand side of (\ref{bill}) is the mean of $N$ identically distributed independent variables, by way of central limit theorem we approximated the distribution of fitnesses using a Gaussian 
distribution.  The quartile boundaries were then determined from this approximation.  Some test simulations showed this to be a reasonably accurate approximation.  

To explore fully the changing nature of epistasis along an adaptive walk, for the initial genotype we sampled from genotypes with fitness below the mean minus 1.5 standard deviations 
according to the theoretical approximation.  This corresponds (again, theoretically) to the $0.067$ quantile of the distribution.

Our Rough Mt.\ Fuji fitness landscapes were constructed in the spirit of their namesakes in the wider literature.  At first, each genotype is assigned a deterministic fitness component given as follows:

\begin{equation*}
\mathit{slope} \cdot \frac{\textit{\# of loci in '1' state}}{N}
\end{equation*}

\noindent where {\it slope} is a pre--determined fixed parameter.  To each of these deterministic values a random value drawn from a uniform distribution on $[0,1]$ is added.  

\begin{equation*}
\mathit{slope} \cdot \frac{\textit{\# of loci in '1' state}}{N} + RN_{genotype} 
\end{equation*}

\noindent Finally, we applied a linear transformation making the minimum and maximum fitnesses $0.05$ and $1$ respectively.  Note that by our construction the ``expected'' fitness 
difference between the genotypes $000\ldots 0$ and $111 \ldots 1$ will be $0.95 \cdot {\it slope}$.  The parameter {\it slope} determined the relative contributions of the deterministic 
component and the noise component in the landscape, with high values of {\it slope} implying a low ratio of noise component to deterministic component.  

Since the computation of empirical quantiles was feasible for Rough Mt.\ Fuji landscapes, we used them for determining quartile boundaries and selecting initial genotypes.  The latter were selected 
from those genotypes with fitnesses among the bottom $0.067$, as they were chosen in the $NK$ landscape case, but in this case using the empirical quantile rather than the theoretical 
quantile.  

All simulations were coded in the programming language R.

\newpage

\section*{Appendix}

\section*{\scriptsize Table Key}

In each diagram, the relative proportions are given for antagonistic, synergistic, and sign epistasis as defined in the main article.  
The proportions represented are among those quadruples which had epistasis.  
Figures 1--8 are the results of the ``stratified'' sampling as described in the main text.  The proportions rerpresented in the 
``All'' column are the result of 10000 simulations.  Each quartile column, labeled Q1 through Q4 is the result of 2500 simulations each.  
In Figures  9--24, 10,000 fitness landscapes are simulated.  In each, an initial genotype is selected 
in the manner described in the appendix.  An adaptive walk is then simulated.  Upon reaching the third genotype in the walk, the epistasis is calculated for that genotype and the previous two.  
This first calculation corresponds to Step 1 in the horizontal axis.  
The relative proportions of subsequent calculations of epistasis are recorded in Step 2, Step 3, etc.  The meaning of the 
parameters used in generating the fitness landscapes is discussed in Materials and Methods.

\newpage

\hspace{-0.1\linewidth}
\begin{minipage}[h]{0.45\linewidth}
\includegraphics[scale=0.55]{bar1.pdf}
\end{minipage}
\hspace{0.15\linewidth}
\begin{minipage}[h]{0.45\linewidth}
\includegraphics[scale=0.55]{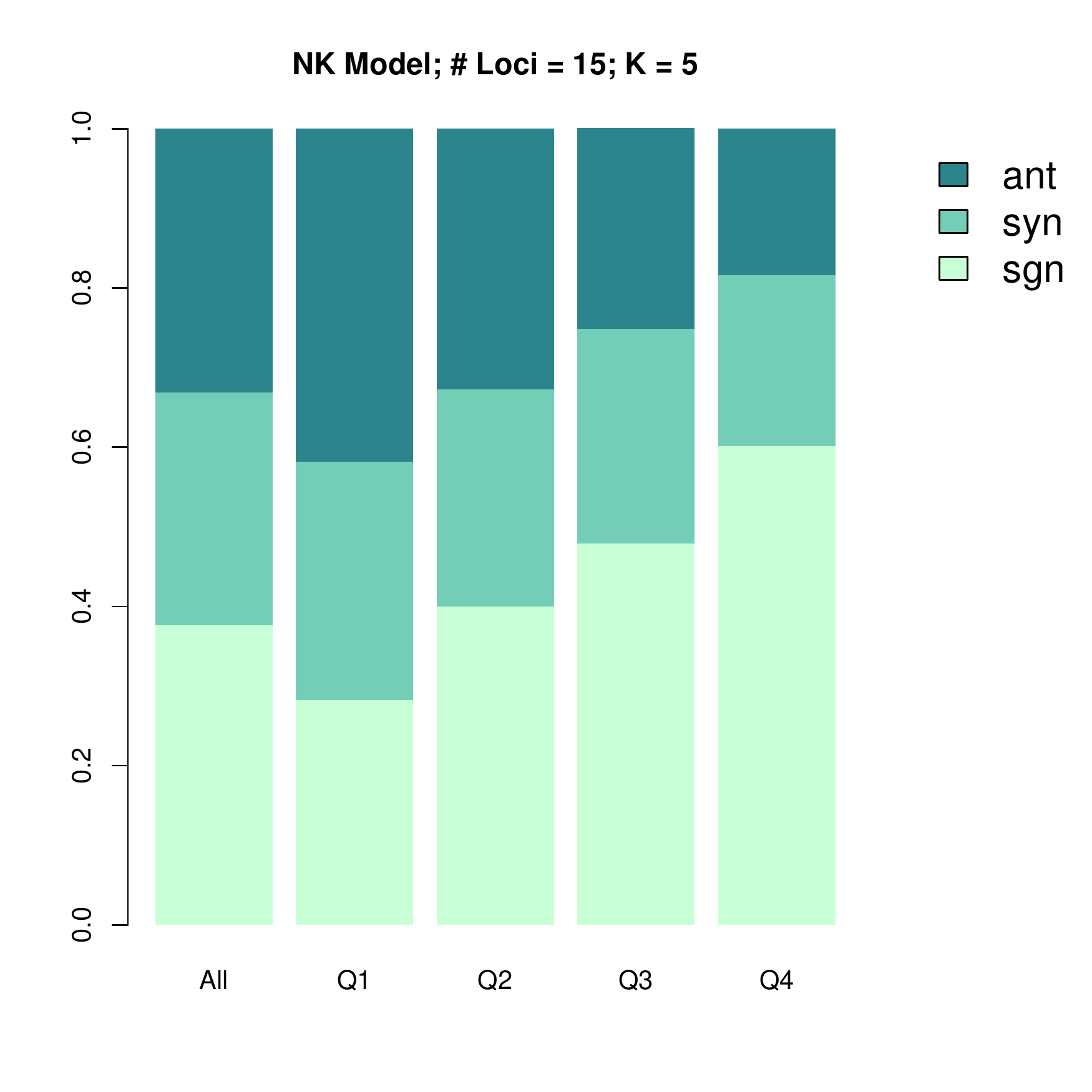}
\end{minipage}

\hspace{-0.1\linewidth}
\begin{minipage}[h]{0.45\linewidth}
\includegraphics[scale=0.55]{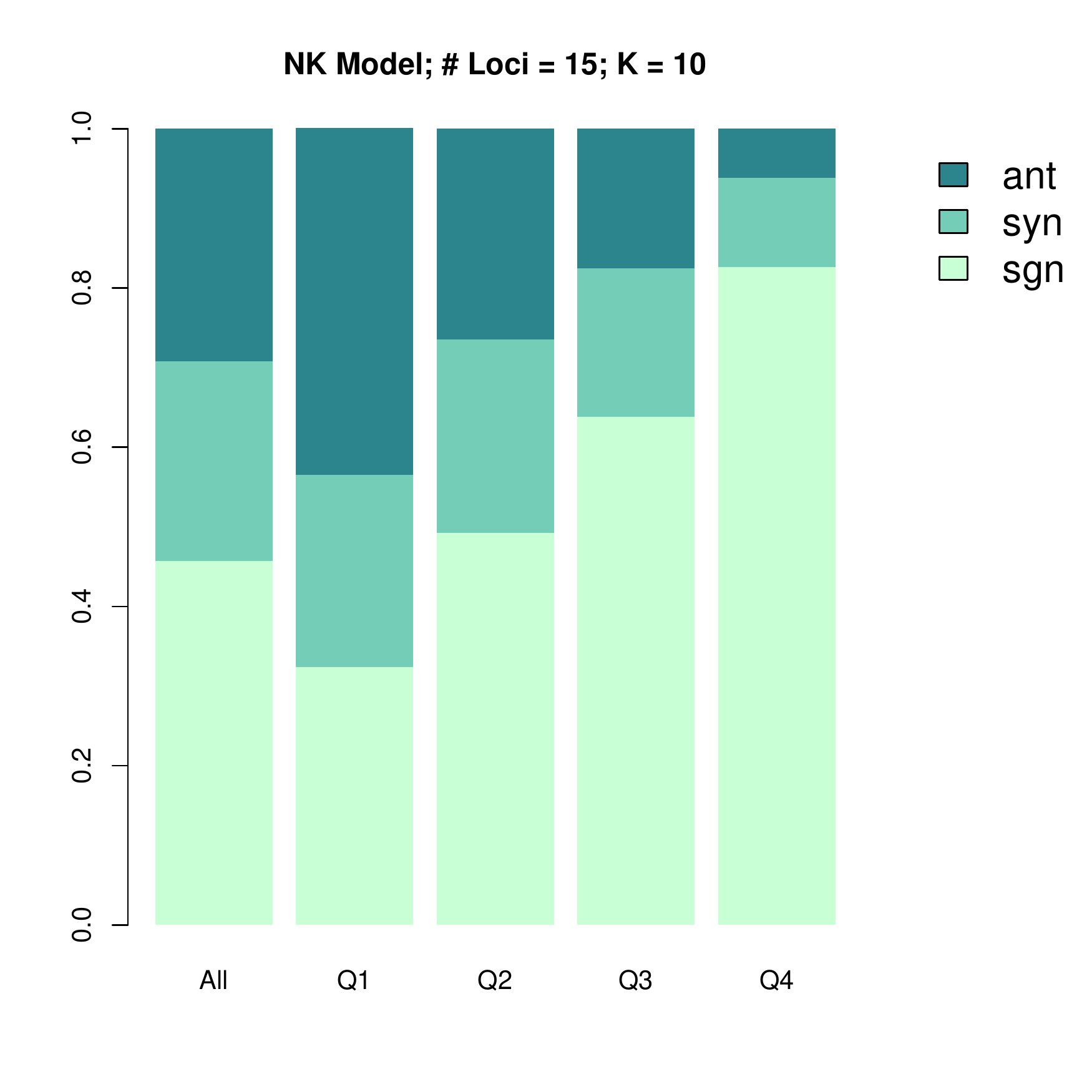}
\end{minipage}
\hspace{0.15\linewidth}
\begin{minipage}[h]{0.45\linewidth}
\includegraphics[scale=0.55]{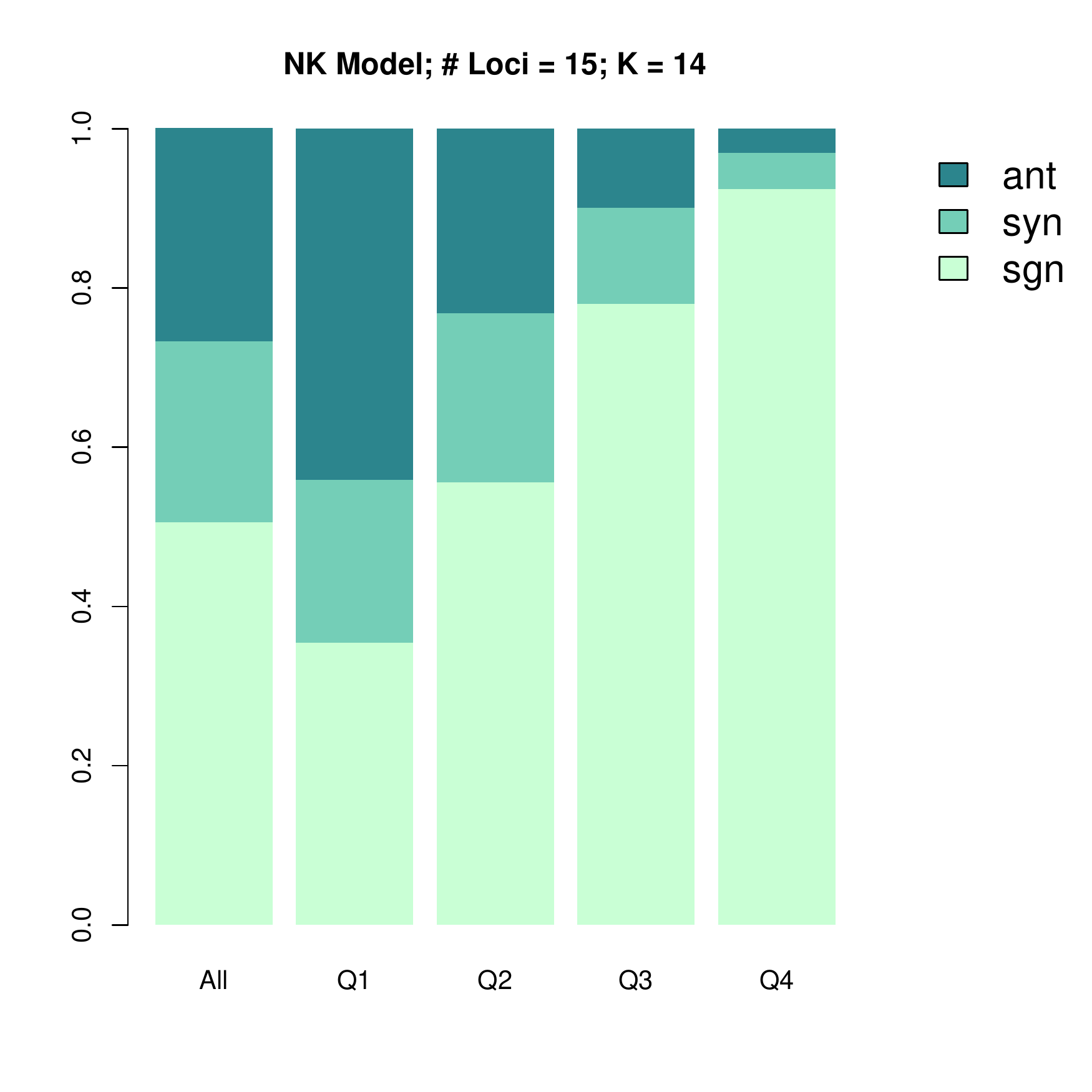}
\end{minipage}

\newpage

\hspace{-0.1\linewidth}
\begin{minipage}[h]{0.45\linewidth}
\includegraphics[scale=0.55]{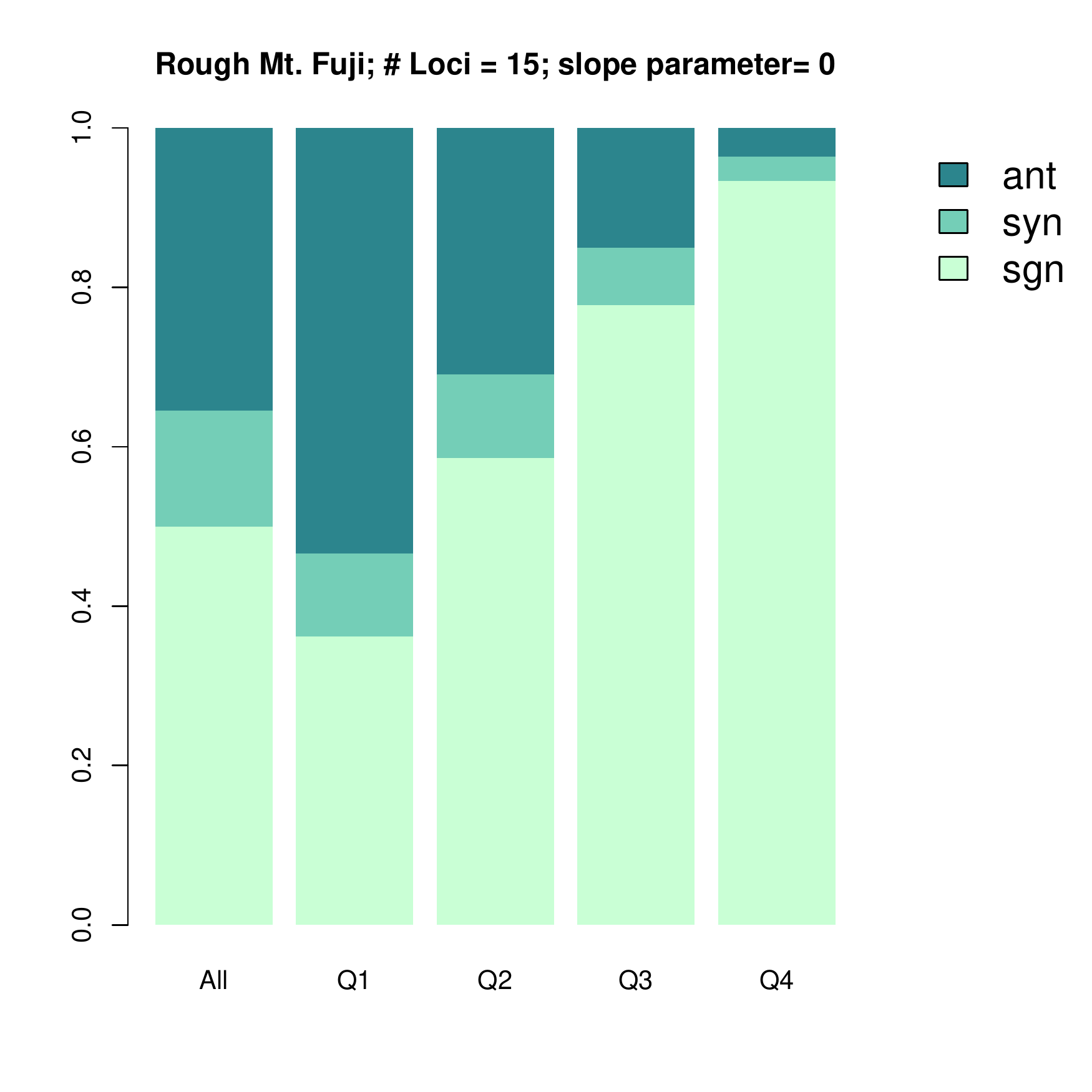}
\end{minipage}
\hspace{0.15\linewidth}
\begin{minipage}[h]{0.45\linewidth}
\includegraphics[scale=0.55]{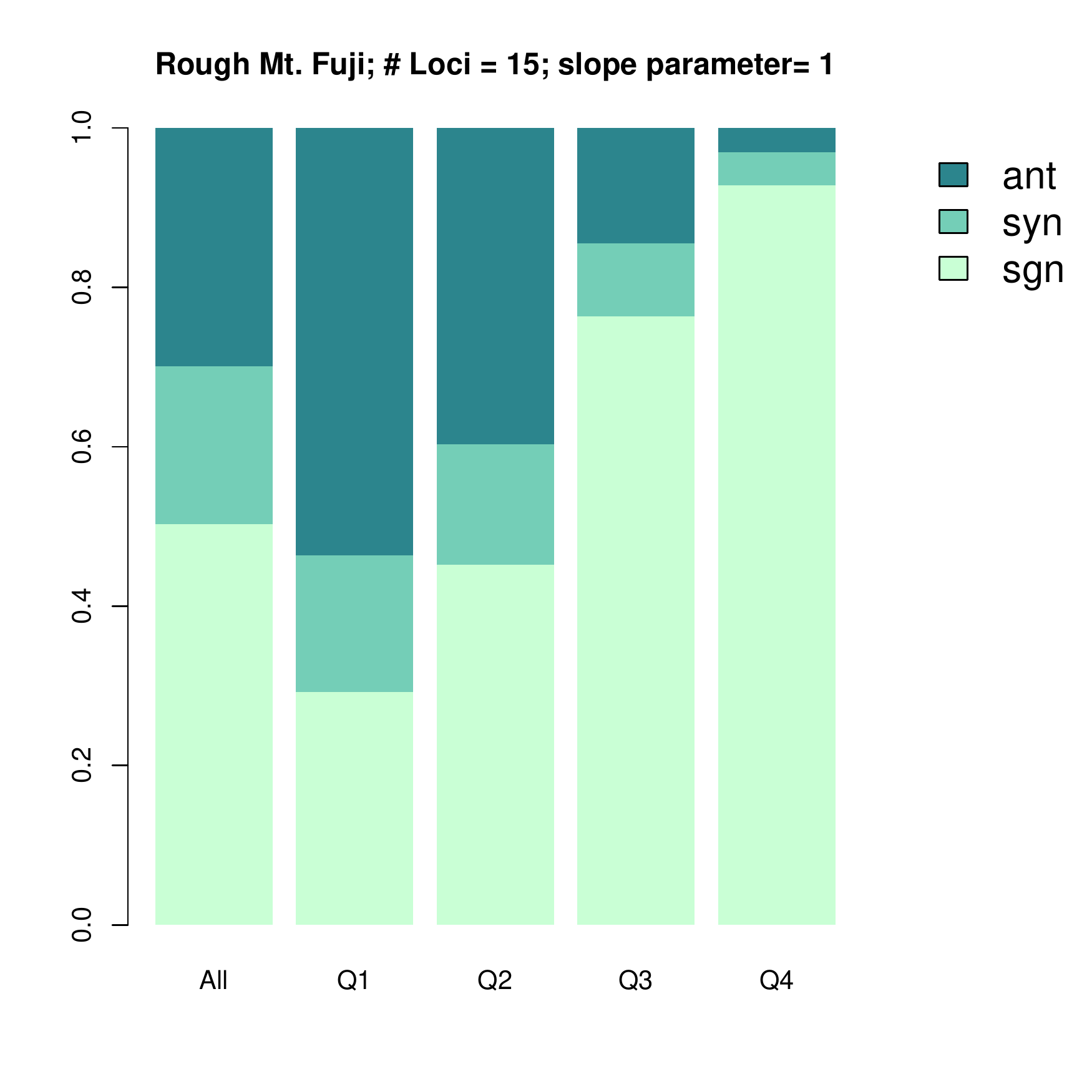}
\end{minipage}

\hspace{-0.1\linewidth}
\begin{minipage}[h]{0.45\linewidth}
\includegraphics[scale=0.55]{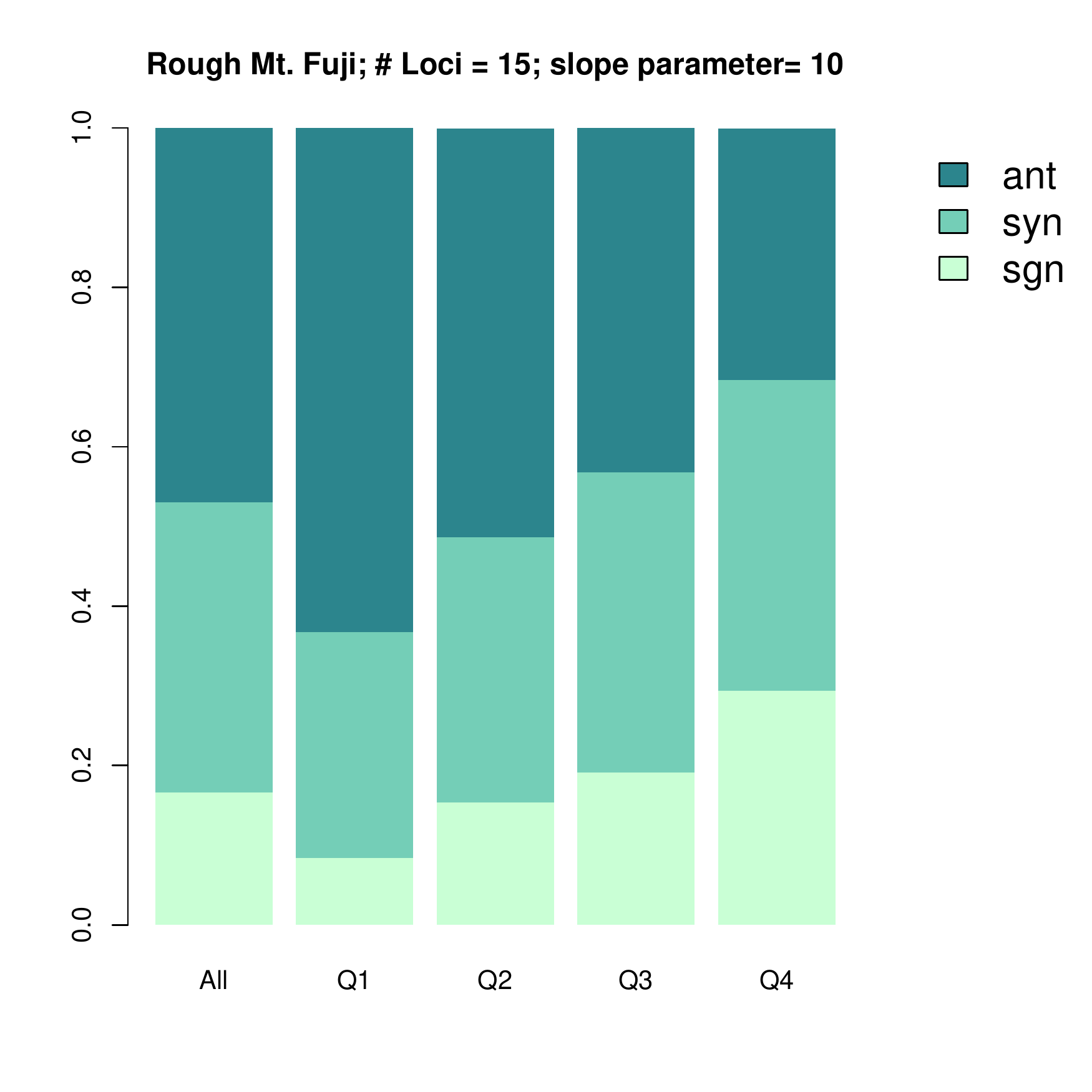}
\end{minipage}
\hspace{0.15\linewidth}
\begin{minipage}[h]{0.45\linewidth}
\includegraphics[scale=0.55]{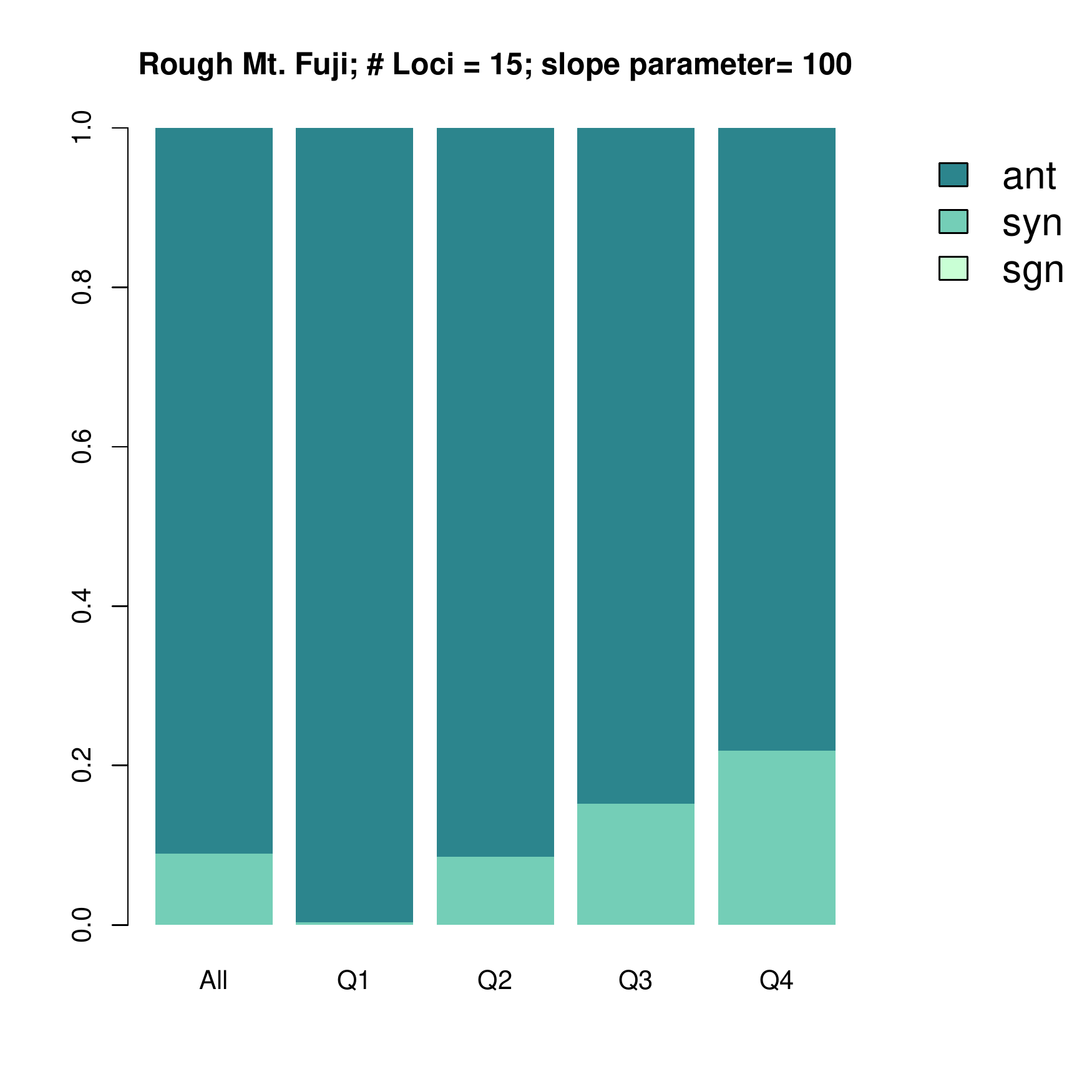}
\end{minipage}

\newpage

\hspace{-0.1\linewidth}
\begin{minipage}[h]{0.45\linewidth}
\includegraphics[scale=0.55]{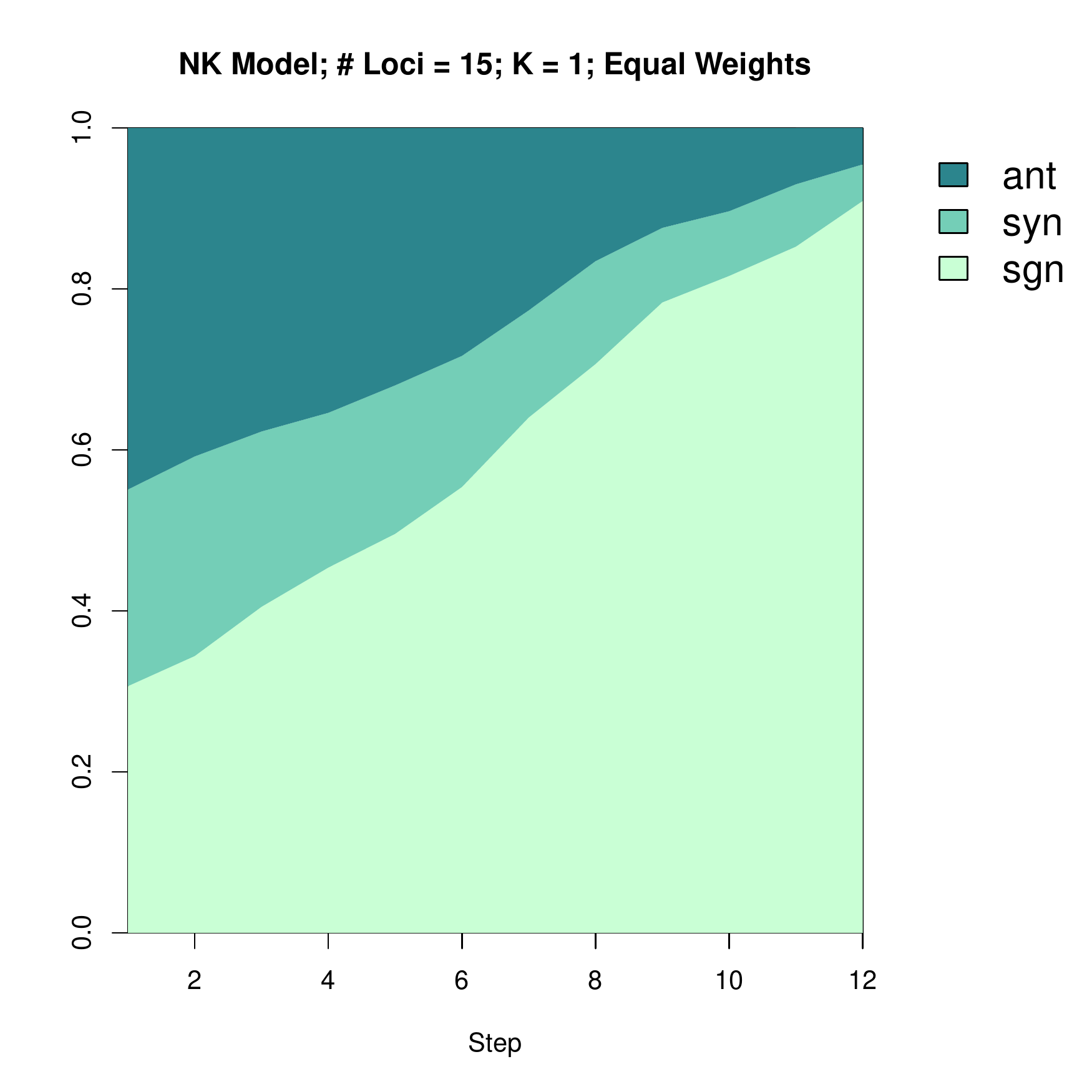}
\end{minipage}
\hspace{0.15\linewidth}
\begin{minipage}[h]{0.45\linewidth}
\includegraphics[scale=0.55]{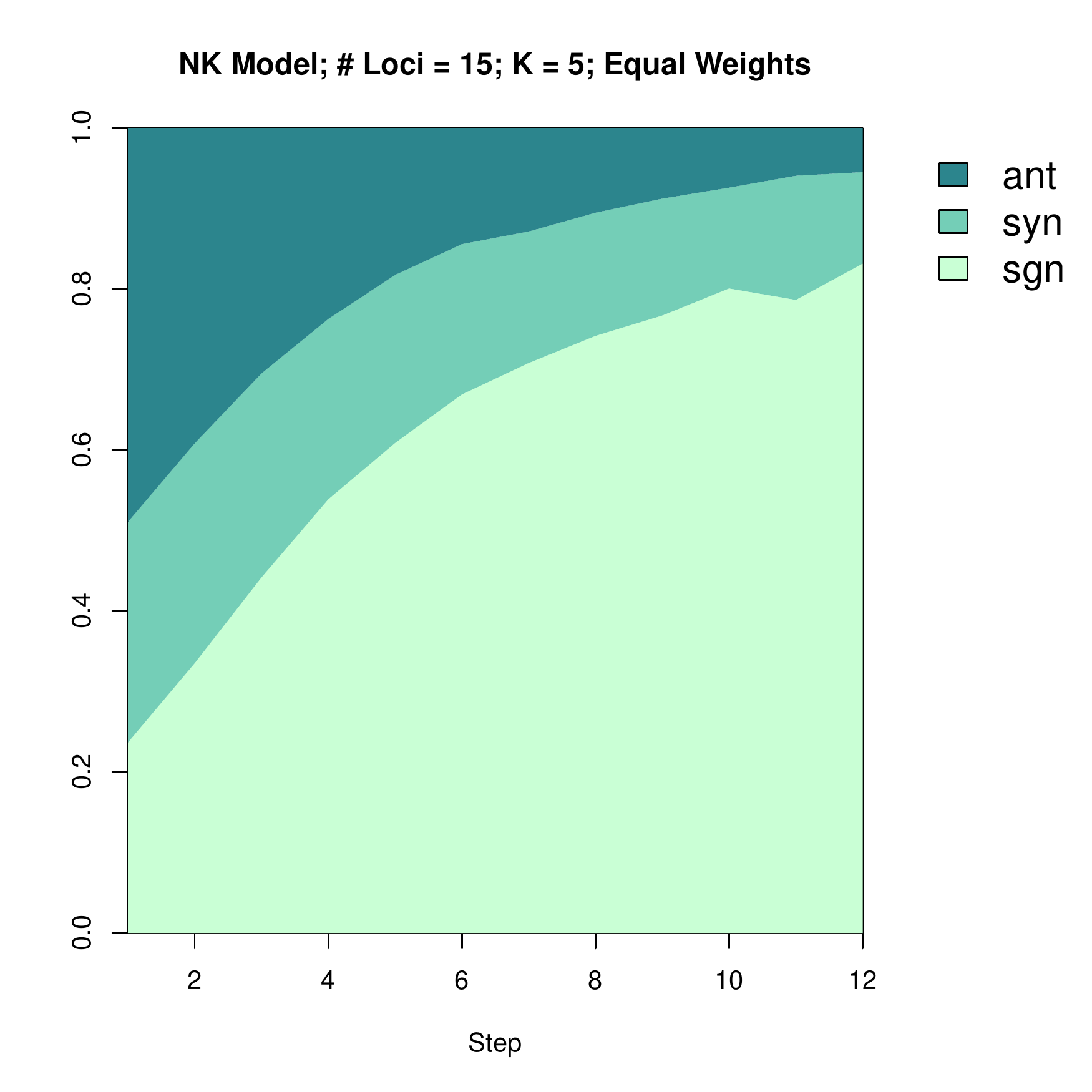}
\end{minipage}

\hspace{-0.1\linewidth}
\begin{minipage}[h]{0.45\linewidth}
\includegraphics[scale=0.55]{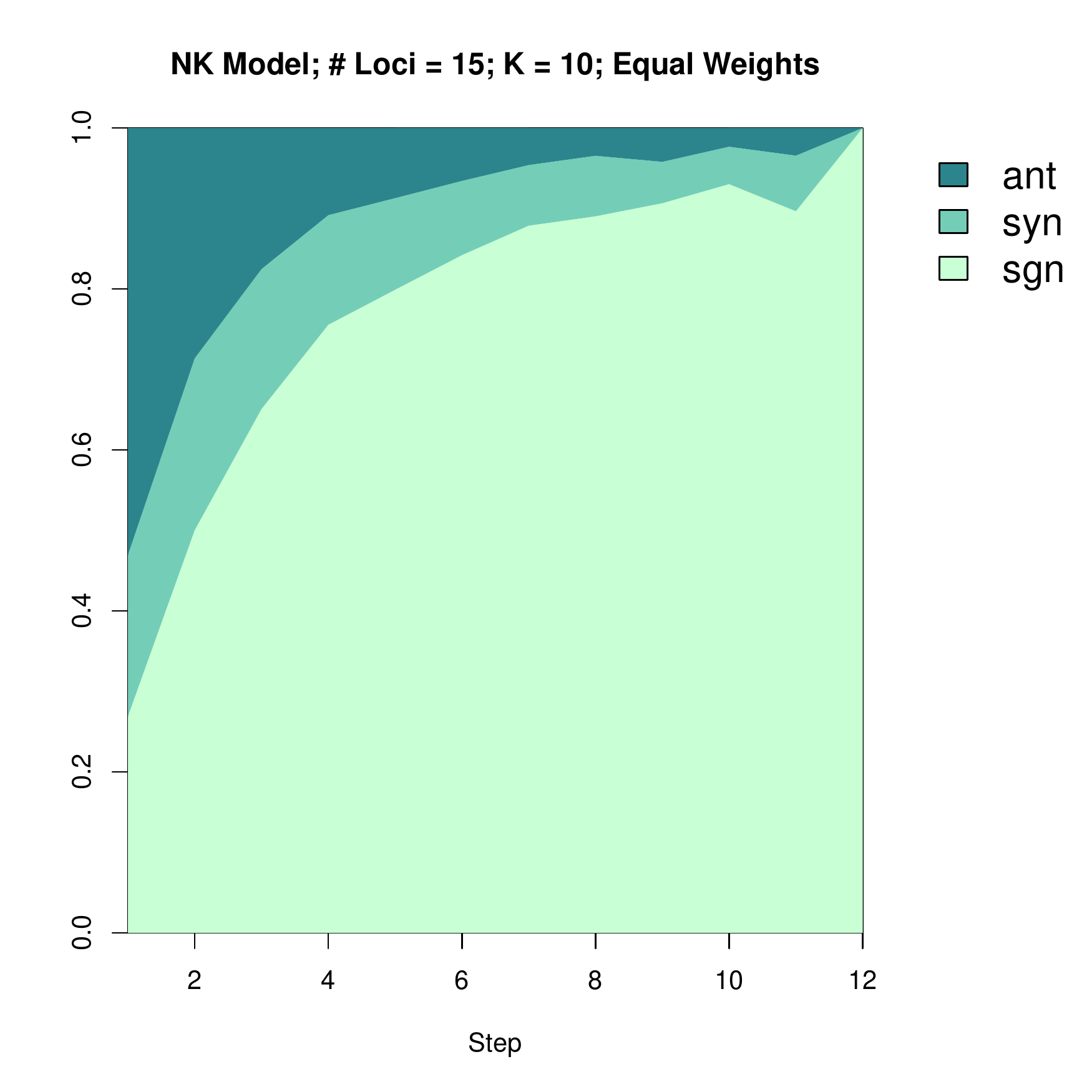}
\end{minipage}
\hspace{0.15\linewidth}
\begin{minipage}[h]{0.45\linewidth}
\includegraphics[scale=0.55]{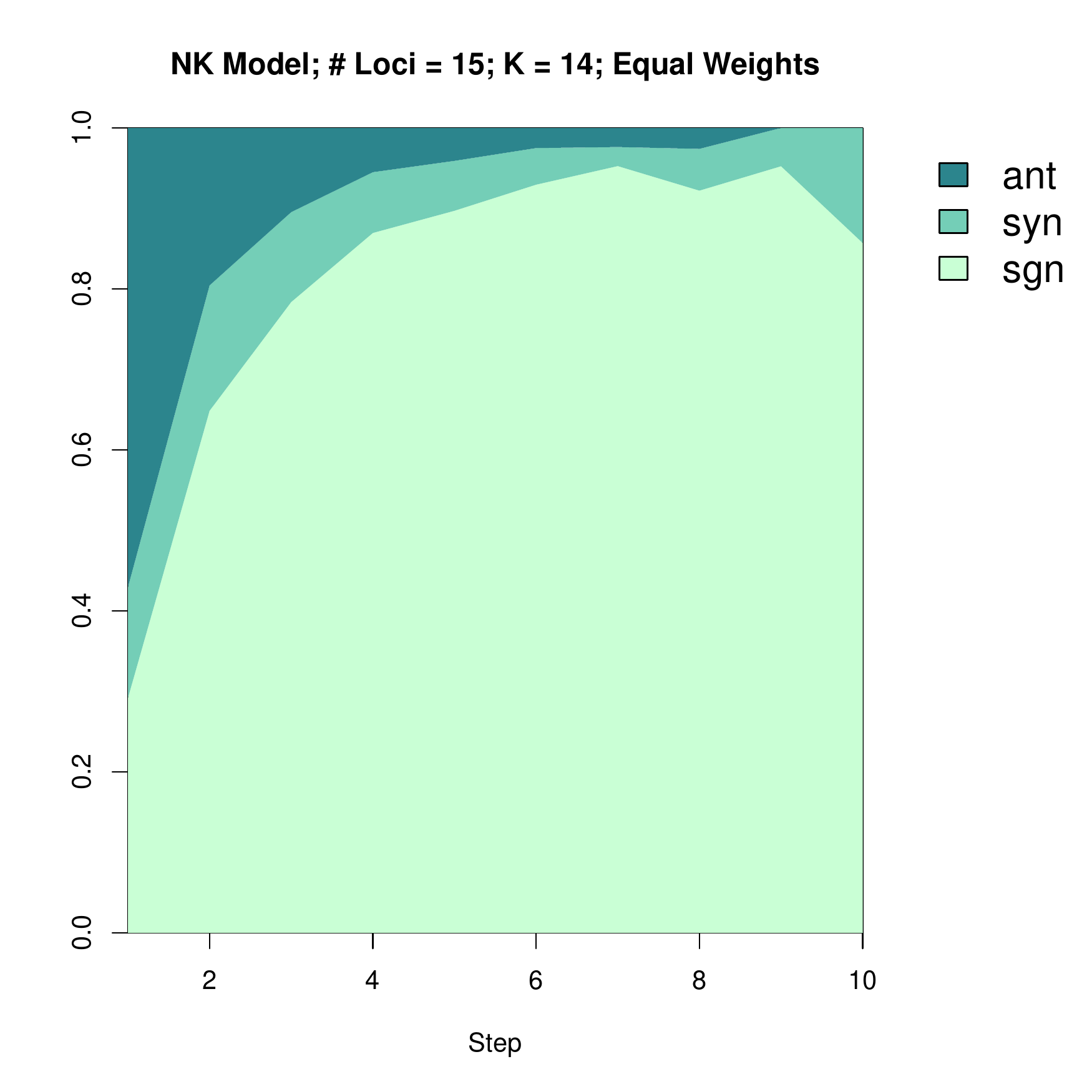}
\end{minipage}

\newpage

\hspace{-0.1\linewidth}
\begin{minipage}[h]{0.45\linewidth}
\includegraphics[scale=0.55]{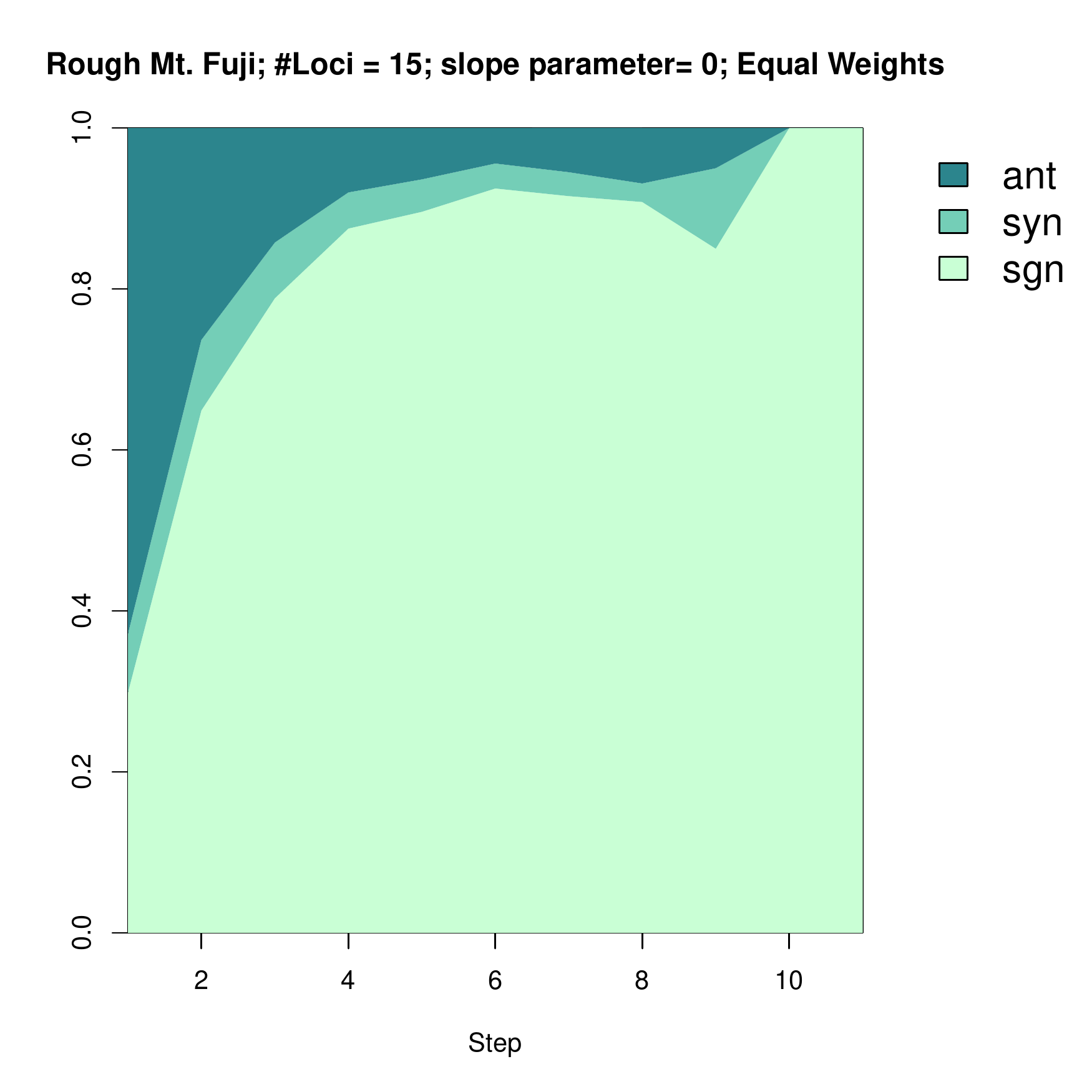}
\end{minipage}
\hspace{0.15\linewidth}
\begin{minipage}[h]{0.45\linewidth}
\includegraphics[scale=0.55]{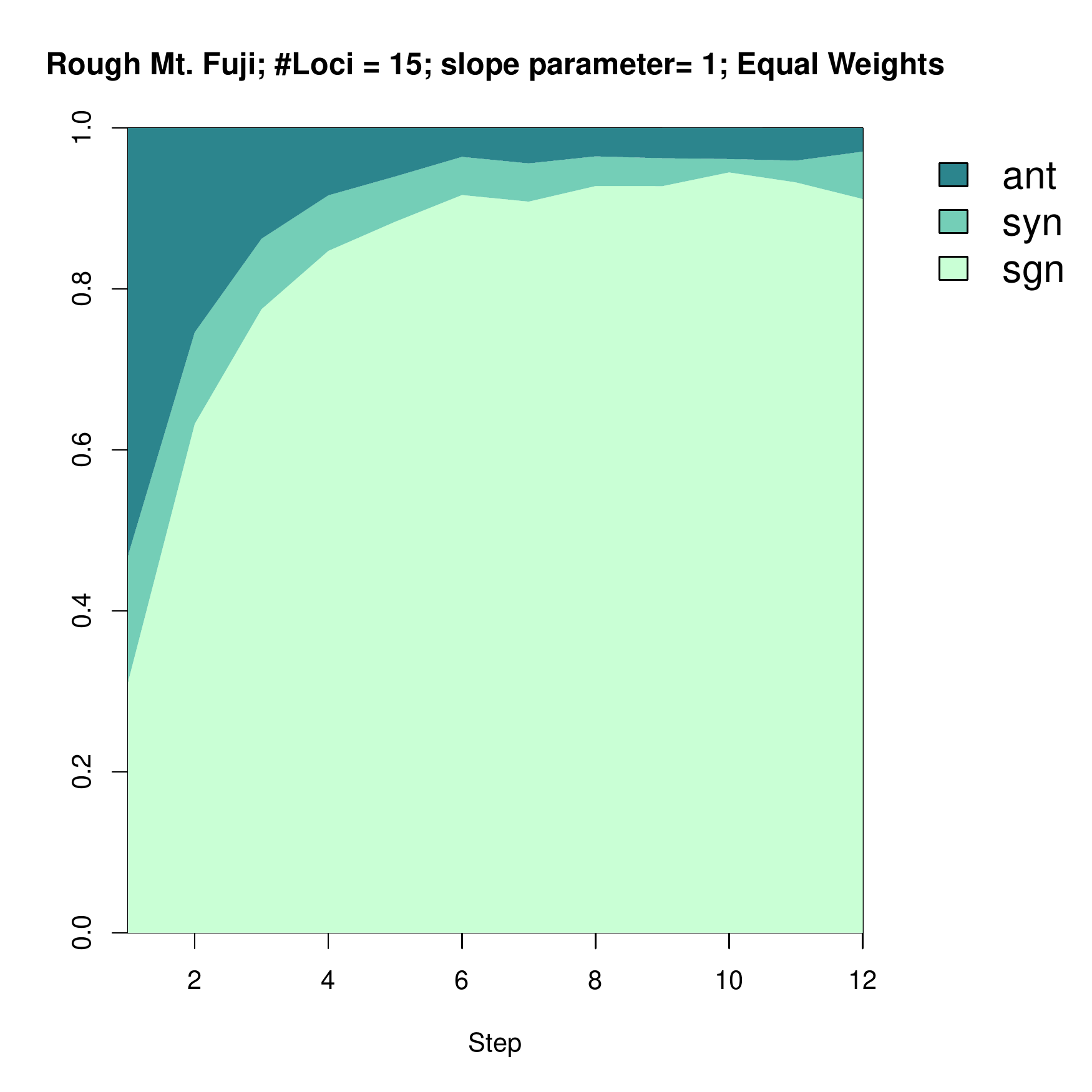}
\end{minipage}

\hspace{-0.1\linewidth}
\begin{minipage}[h]{0.45\linewidth}
\includegraphics[scale=0.55]{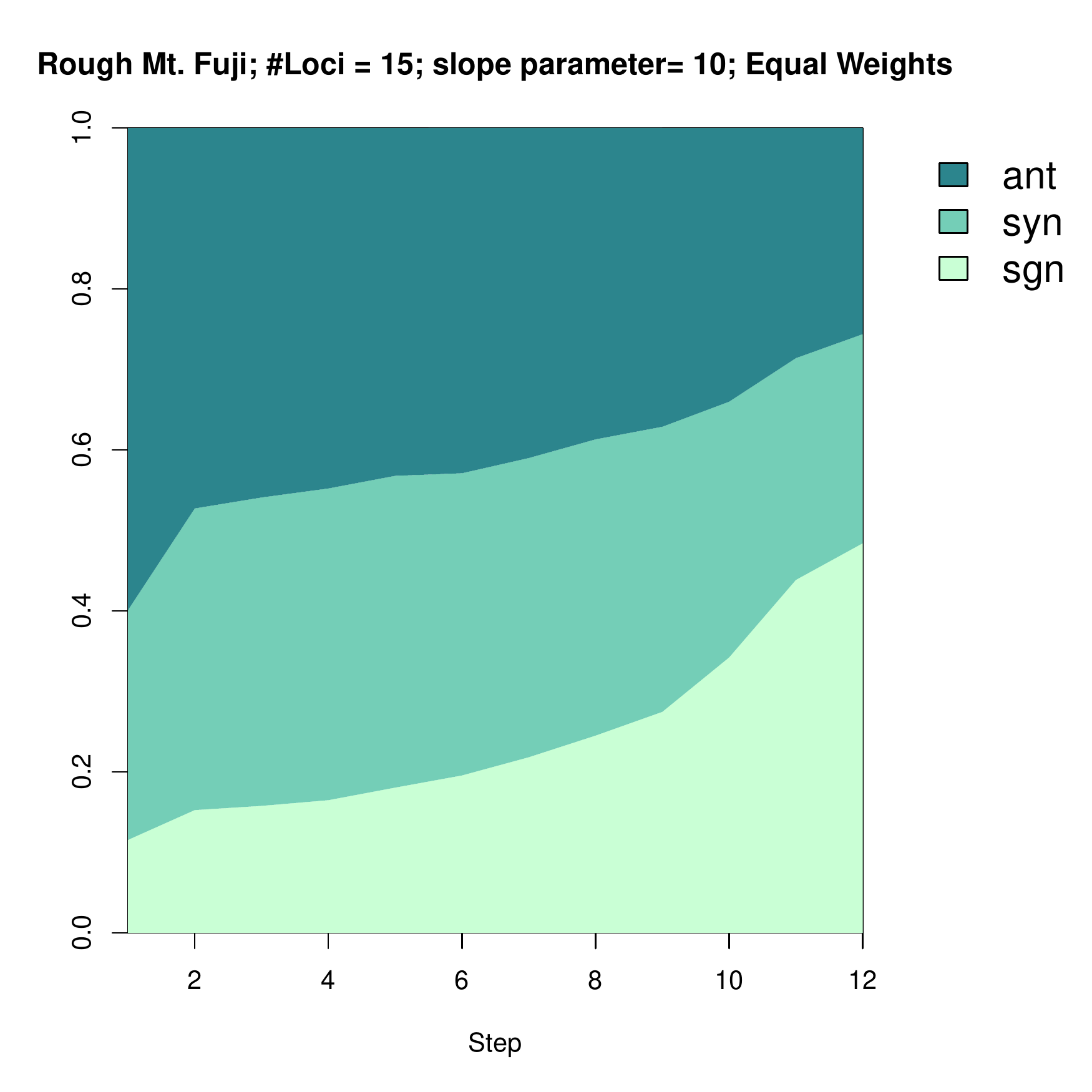}
\end{minipage}
\hspace{0.15\linewidth}
\begin{minipage}[h]{0.45\linewidth}
\includegraphics[scale=0.55]{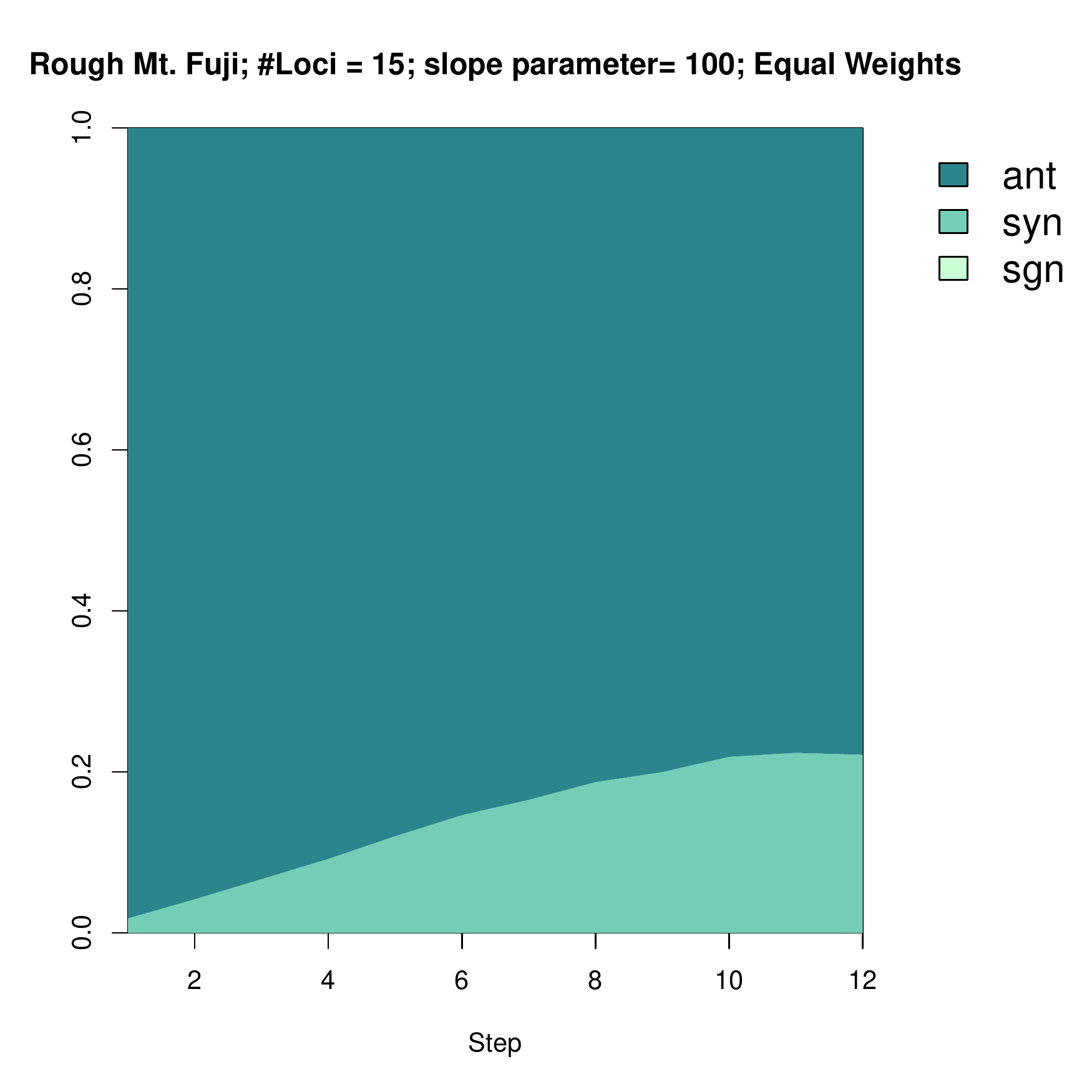}
\end{minipage}

\newpage

\hspace{-0.1\linewidth}
\begin{minipage}[h]{0.45\linewidth}
\includegraphics[scale=0.55]{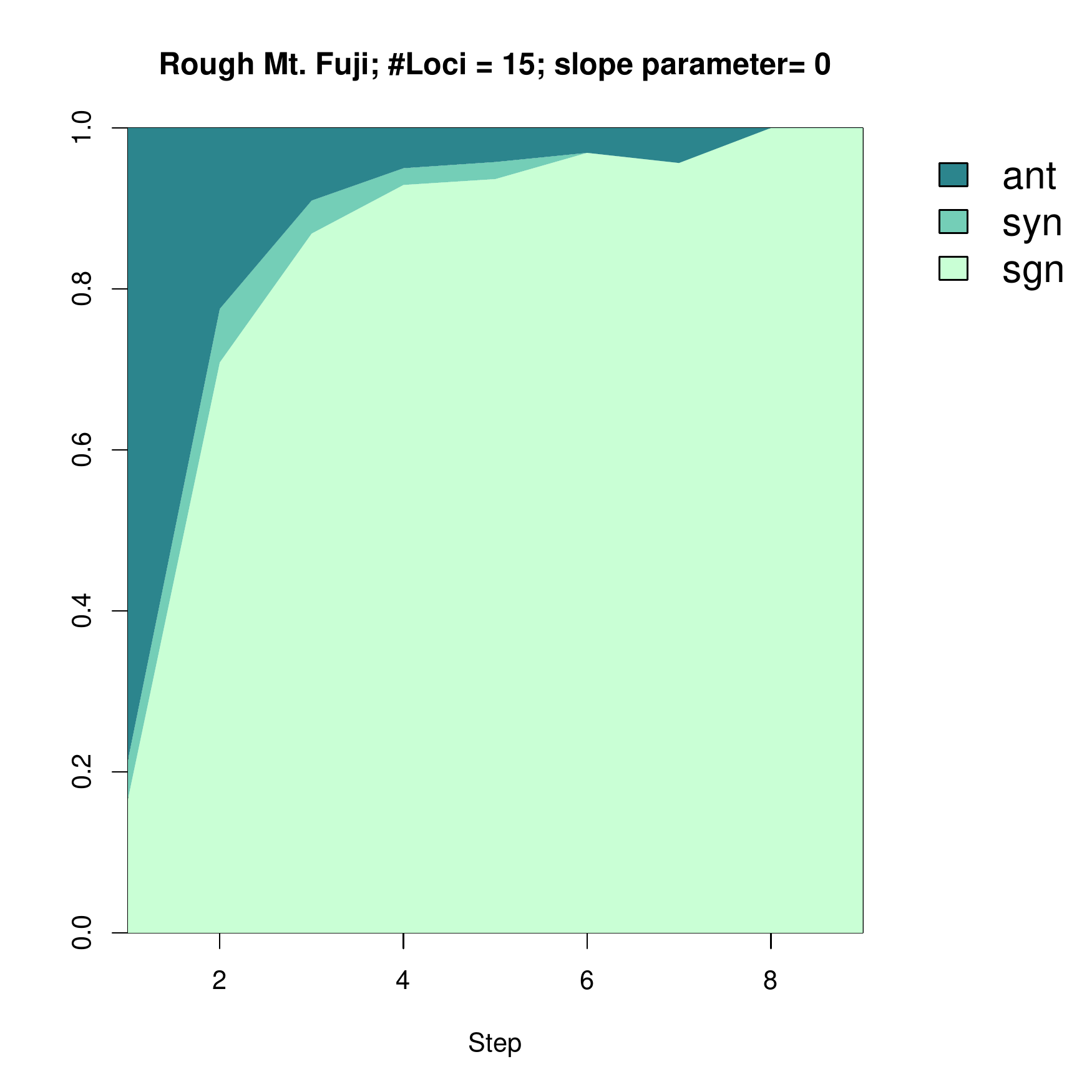}
\end{minipage}
\hspace{0.15\linewidth}
\begin{minipage}[h]{0.45\linewidth}
\includegraphics[scale=0.55]{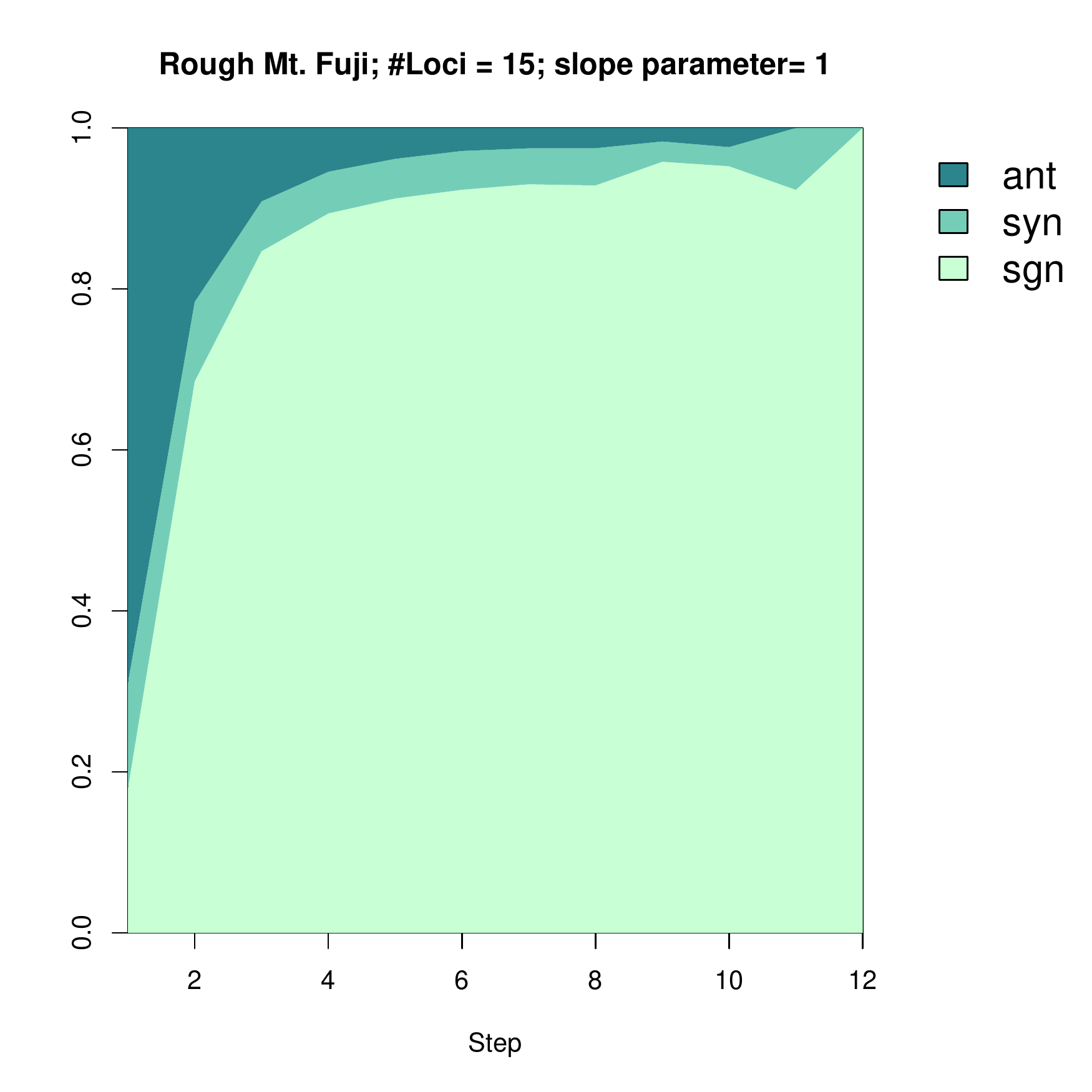}
\end{minipage}

\hspace{-0.1\linewidth}
\begin{minipage}[h]{0.45\linewidth}
\includegraphics[scale=0.55]{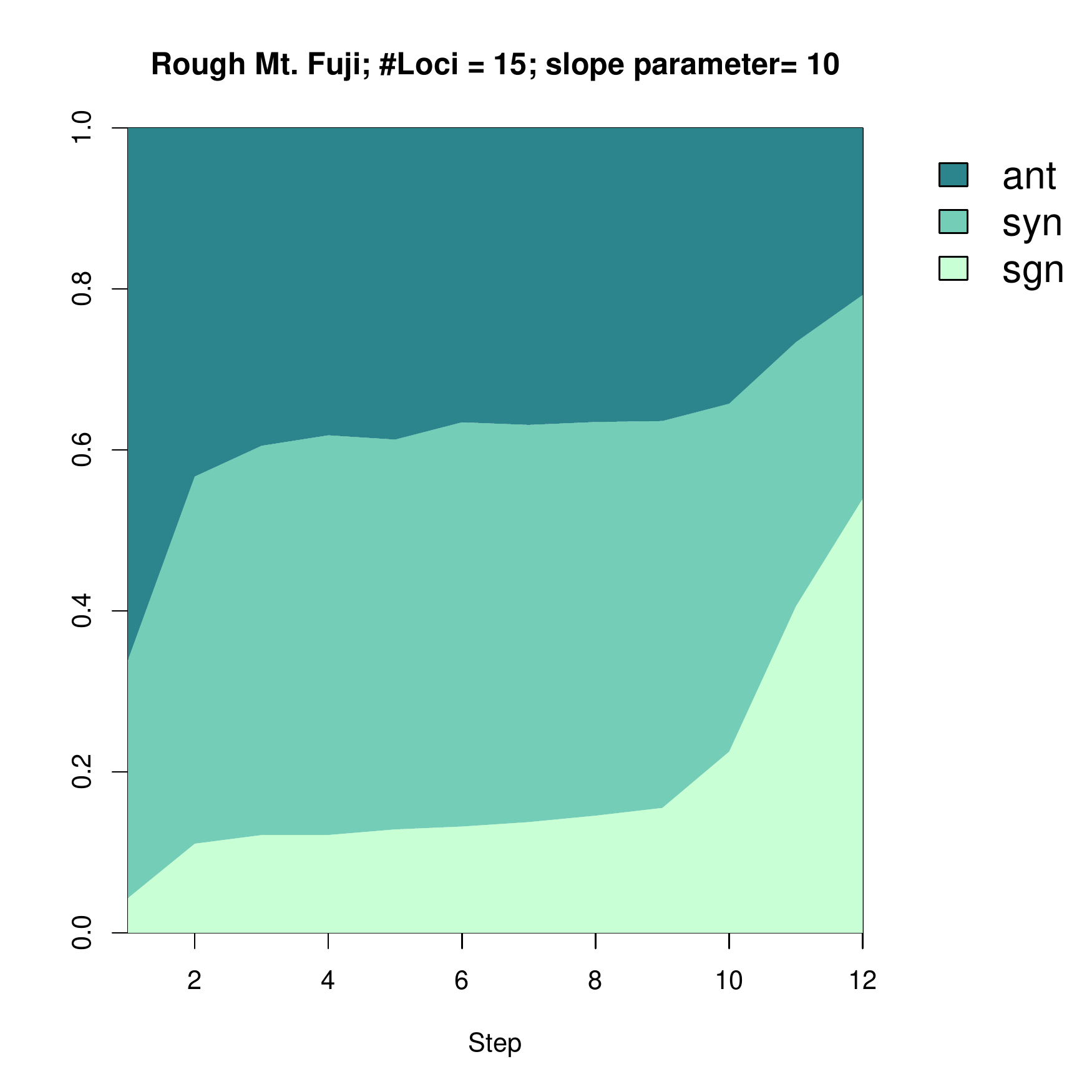}
\end{minipage}
\hspace{0.15\linewidth}
\begin{minipage}[h]{0.45\linewidth}
\includegraphics[scale=0.55]{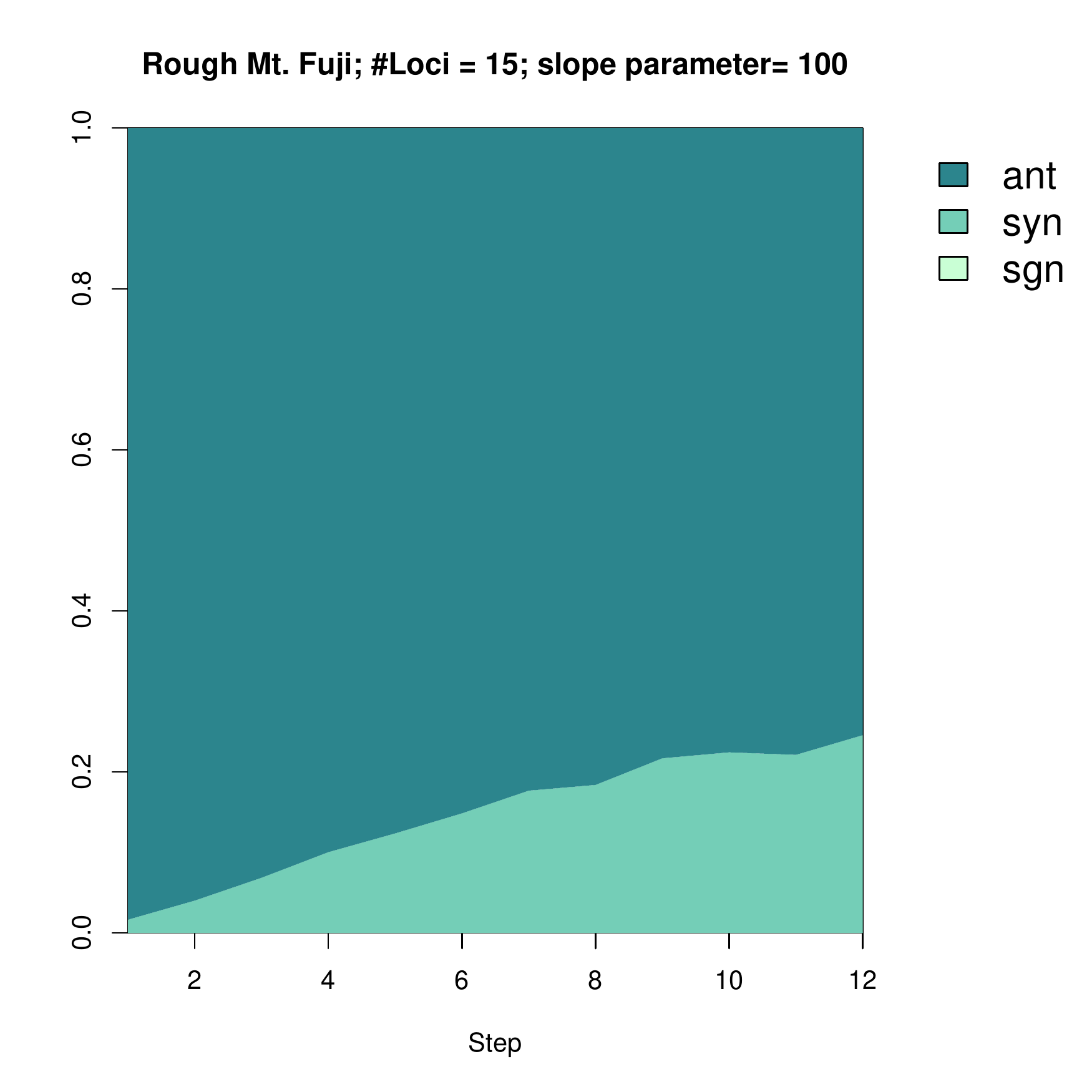}
\end{minipage}

\newpage

\hspace{-0.1\linewidth}
\begin{minipage}[h]{0.45\linewidth}
\includegraphics[scale=0.55]{graph13.pdf}
\end{minipage}
\hspace{0.15\linewidth}
\begin{minipage}[h]{0.45\linewidth}
\includegraphics[scale=0.55]{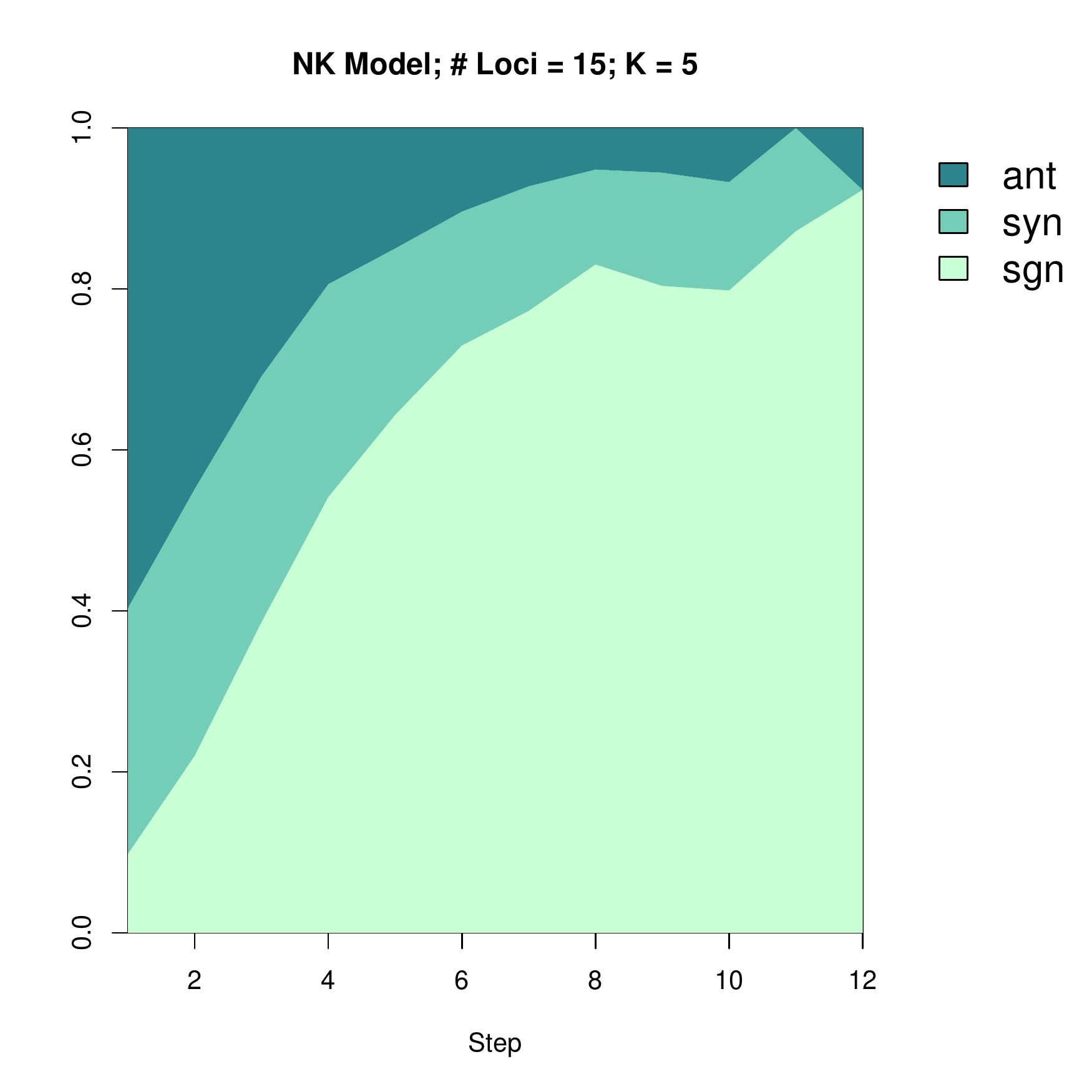}
\end{minipage}

\hspace{-0.1\linewidth}
\begin{minipage}[h]{0.45\linewidth}
\includegraphics[scale=0.55]{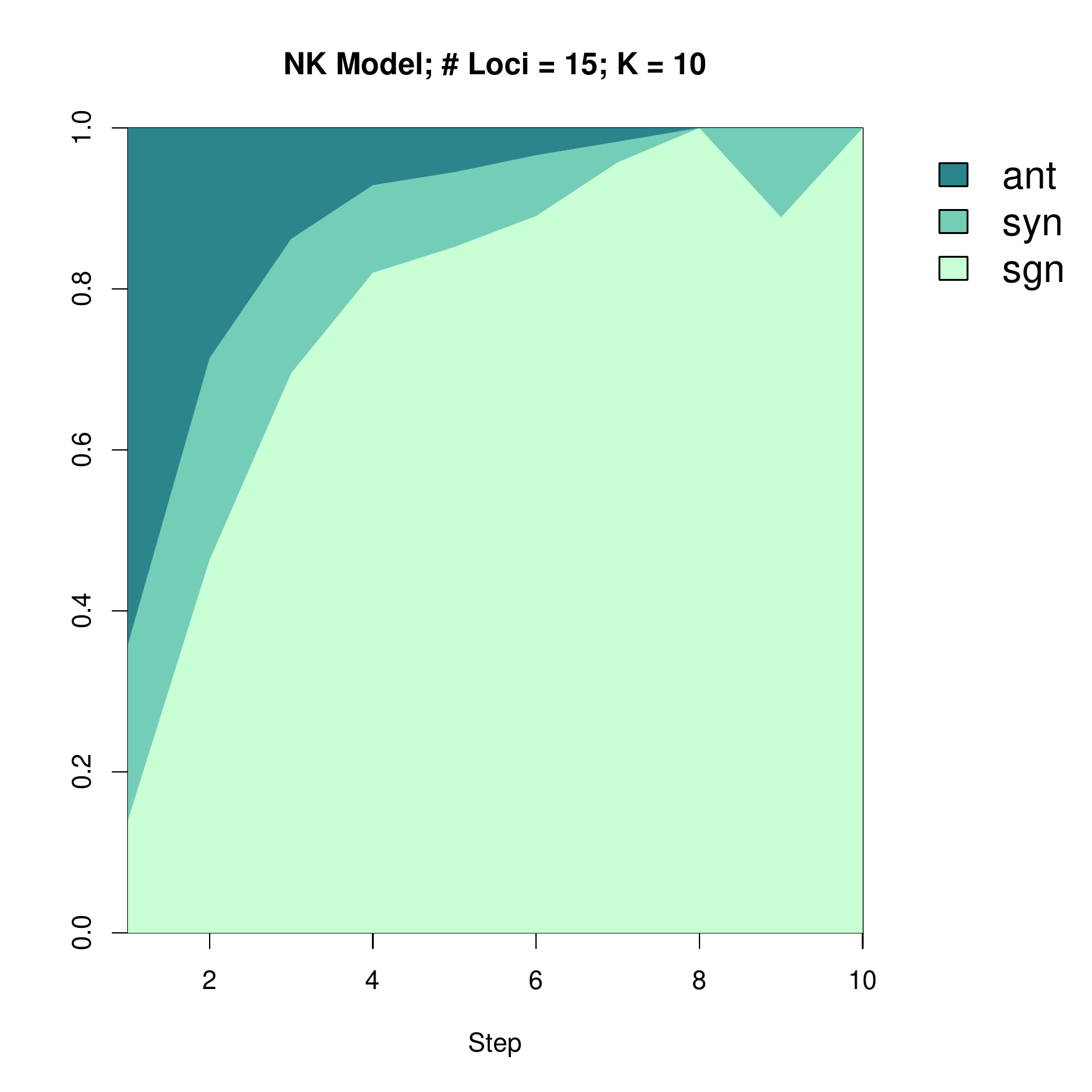}
\end{minipage}
\hspace{0.15\linewidth}
\begin{minipage}[h]{0.45\linewidth}
\includegraphics[scale=0.55]{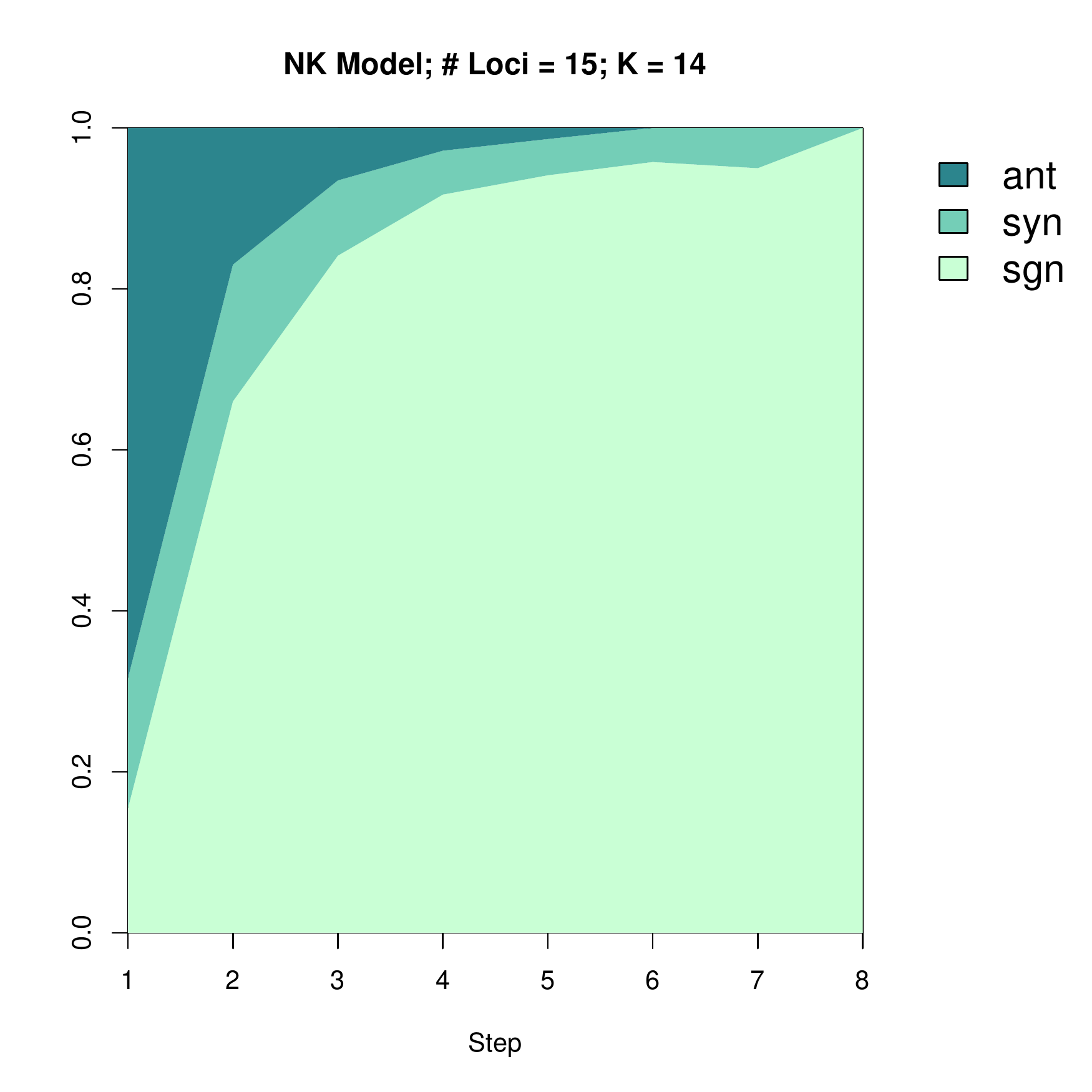}
\end{minipage}


\begin{thebibliography}{99}

\bibitem[Aita and Husimi, 1996]{ah}
Aita, T., Husimi, Y. (1996).
Fitness spectrum among random mutants on 
Rough Mt. Fuji-type fitness landscape.
{\emph{J Theor Biol.}}182(4):469-85.

\bibitem[Aita et al., 2000]{aui}
Aita, T., Uchiyama, H., Inaoka, T., Nakajima, M., Kokubo, T. and Husimi, Y.
(2000).
Analysis of a local fitness landscape with a model of the rough Mt. 
Fuji-type landscape: application to prolyl endopeptidase and thermolysin.
{\emph{Biopolymers}}54(1): 64-79.
 


\bibitem[Beerenwinkel et al., 2007 a]{bps}
Beerenwinkel, N., Pachter, L. and Sturmfels, B. (2007).
Epistasis and shapes of fitness landscapes.
{\emph{Statistica Sinica}} 17:1317--1342.




\bibitem[Beerenwinkel et al., 2007 b]{bpse}
Beerenwinkel, N., Pachter, L., Sturmfels, B., Elena, S. F. and
Lenski, R. E. (2007). Analysis of epistatic interactions and fitness
landscapes using a new geometric approach.
{\emph{BMC Evolutionary Biology}} 7:60.



\bibitem[Chou et al., 2011]{ccd}
Chou, H.H., Chiu, H.C., Delaney, N.F., Segre, D., Marx, C.J. (2011).
Diminishing returns 
epistasis among beneficial mutations decelerates adaptation. 
{\emph{Science}} 
332:1190--1192.


\bibitem[Crona, 2013]{c}
Crona, K. (2013) Graphs, polytopes and fitness landscapes (book Chapter), 
Springer, Chapter 7, 177-206.
Recent Advances in the Theory and Application of Fitness Landscapes 
(A. Engelbrecht and H. Richter, eds.). Springer Series in Emergence, 
Complexity, and Computation, 2013.
Springer, Chapter 7, 177-206.

\bibitem[Crona et al., 2013]{kelly}
Crona, K., Patterson, D., Stack, K.,
Greene, D.,  Goulart, C.,
Mahmudi, M., Jacobs, S. D., Kallman, M.
and Barlow, M.
(2013)
A quantification of theory-data incompatibility for fitness landscapes.
{\tt arXiv:1303.3842}

\bibitem[Crona et al., 2013]{cgb}
Crona, K., Greene, D. and Barlow, M. (2013).
The peaks and geometry of fitness landscapes.
{\emph{J. Theor. Biol.}} 317: 1--13.






\bibitem[De Visser et al., 2009]{dpk}
De Visser, J. A. G. M., Park, S.C. and Krug, J.( 2009).
 Exploring the effect of sex on empirical fitness landscapes.
{\emph{The American Naturalist.}}



\bibitem[Draghi and Plotkin, 2013]{dp}
Draghi, J. A. and Plotkin, J. B. (2013).
Selection biases the prevalence and type of epistasis along adaptive trajectories. 
{\emph{Evolution}} doi: 10.1111/evo.12192.

\bibitem[Flyvbjerg and Lautrup, 1992]{fl}
Flyvbjerg, H. and Lautrup, B. (1992)
Evolution in a rugged fitness landscape
{\emph{Physical Review A}} 6714-6723.

\bibitem[Franke et al., 2011]{fkd}
Franke, J., Kl{\"o}zer, A., de Visser, J.A.G.M.
and Krug., J. (2011). Evolutionary Accessibility
of Mutational Pathways.
{\emph{PLoS Comput Biol}} 7(8):
e1002134. doi:10.1371/journal.pcbi.1002134.


\bibitem[Gilliespie, 1983]{g1983}
Gillespie, J. H. (1983). A simple stochastic gene substitution model.
{\emph{Theor. Pop. Biol.}} 23 : 202--215.

\bibitem[Gilliespie, 1984]{g1984}
Gillespie, J. H. (1984). The molecular clock may be an episodic
clock. {\emph{Proc. Natl. Acad. Sci. USA 81}} : 8009--8013.


\bibitem[Goulart et al., 2013]{gmc}
Goulart, C. P., Mentar, M., Crona, K., Jacobs, S. J., Kallmann, M.,
Hall, B. G., Greene D., Barlow M. (2013). Designing antibiotic
cycling strategies by determining and understanding local
adaptive landscapes.
{\emph{PLoS ONE}} 8(2): e56040.
doi:10.1371/journal.pone.0056040.






\bibitem[Kauffman and Levin, 1987]{kl}
Kauffman, S. A. and Levin, S. (1987).
Towards a general theory of adaptive
walks on rugged landscapes.
{\emph{J. Theor. Biol}}
128:11--45.

\bibitem[Kauffman and Weinberger, 1989]{kw}
Kauffman, S. A. and Weinberger, E.D. (1989).
The NK model of rugged fitness landscape 
and its application to maturation of the immune response.
{\emph{J. Theor. Biol.}};141:211--245.

\bibitem[Khan et al., 2011]{kds}
Khan AI, Dinh DM, Schneider D, Lenski RE, Cooper TF (2011) Negative 
epistasis between beneficial mutations in an evolving bacterial 
population. Science 332:1193--1196.




%















\bibitem[Macken and Perelson, 1989]{mp}
Macken, C. and Perelson, A. S. (1989).
Protein Evolution on Rugged Landscapes.
{\emph{Proc Natl Acad Sci U S A.}} 
6191-6195.


\bibitem[Orr, 2002]{o}
Orr, H. A. (2001)
The population genetics of adaptation: the adaptation of DNA 
sequences.{\emph{Evolution}} 56:1317-1330.


\bibitem[Maynard Smith, 1970]{s}
Maynard Smith, J. (1970). Natural selection and the concept of protein
space. {\emph{Nature}} 225:563--64.








\bibitem[Poelwijk et al., 2007]{pkw}
Poelwijk, F.J., Kiviet, D. J., Weinreich, D. M. and Tans, S.J. (2007).
Empirical fitness landscapes reveal accessible evolutionary paths.
{\emph{Nature}} 445:383--386.


\bibitem[Poelwijk et al., 2011]{psk}
Poelwijk, F. J., Sorin, T.-N., Kiviet, D. J. and Tans, S. J. (2011).
Reciprocal sign epistasis is a necessary condition for
multi-peaked fitness landscapes.
{\emph{J. Theor. Biol.}} Mar 7; 272(1):141--4. 

\bibitem[R Core Team, 2013]{rsp}
R Core Team (2013). 
R: A Language and Environment for Statistical Computing. 
R Foundation for Statistical Computing, Vienna, Austria.
URL: http://www.R-project.org/.








\bibitem[Karline Soetaert (2013)]{dia}
Soetart, Karline (2013).
diagram: Functions for visualising simple graphs (networks), plotting flow diagrams. R package version 1.6.1.
URL: http://CRAN.R-project.org/package=diagram.

\bibitem[Szendro et al., 2012]{ssf}
Szendro, I. G., Schenk, M. F., Franke, J.
Krug, J. and de Visser J. A. G. M. (2013).
Quantitative analyses of empirical fitness landscapes
{\emph{J. Stat. Mech. P01005.}}





\bibitem[Weinreich et al., 2005] {wwc}
Weinreich, D. M., Watson R. A. and Chao, L. (2005).
Sign epistasis and genetic constraint on evolutionary
trajectories. {\emph{Evolution}} 59, 1165--1174.

\bibitem[Wright, 1931]{w}
Wright, S. (1931). Evolution in Mendelian populations.
{\emph{Genetics}}, 16 97--159.


\end{thebibliography}
\end{document}